\documentclass[10pt]{article}      

\usepackage[british]{babel}
\usepackage{amsmath,amssymb,amsfonts,amsthm,amscd} 
\usepackage{graphics}                 
\usepackage{color}                    
\usepackage{mdframed}                 
\usepackage{mathrsfs}
\usepackage{dsfont}
\usepackage{bbold}
\usepackage{url}         
\usepackage{colonequals} 

\usepackage{cancel}                                               

\usepackage{longtable} 

\usepackage{enumerate}  

\newcommand\Item[1][]{%
  \ifx\relax#1\relax  \item \else \item[#1] \fi
  \abovedisplayskip=0pt\abovedisplayshortskip=0pt~\vspace*{-\baselineskip}}

\makeatletter
  \newcommand{\eqnum}{\leavevmode\hfill\refstepcounter{equation}\textup{\tagform@{\theequation}}} 
\makeatother

\usepackage{nccmath}   
\usepackage[hidelinks]{hyperref}

\usepackage{todonotes}

\usepackage{tikz}
\usetikzlibrary{arrows,decorations.markings,patterns}

\theoremstyle{plain}
\newtheorem{theorem}{Theorem}[section]
\newtheorem{lemma}[theorem]{Lemma}
\newtheorem{proposition}[theorem]{Proposition}
 \newtheorem{corollary}[theorem]{Corollary}
\newtheorem{formalcorollary}[theorem]{Formal corollary}
\theoremstyle{definition}
\newtheorem{definition}[theorem]{Definition}
\theoremstyle{remark}
\newtheorem{remark}[theorem]{Remark}
\AtEndEnvironment{remark}{\qed}

\newcommand{\grad}{\nabla}

\newcommand{\sym}{\mathrm{sym}}
\newcommand{\asym}{\mathrm{asym}}
\newcommand{\drift}{\mathrm{drift}}
\newcommand{\DB}{\mathrm{DB}}

\newcommand{\MFT}{\mathrm{MFT}}
\newcommand{\GEN}{\mathrm{GEN}}

\def\tp{^\mathsf{T}}

\def\Res{\mathrm{Res}}
\def\Det{\mathrm{Det}}
\def\ZR{\mathrm{ZR}}

\DeclareMathOperator*{\argmin}{arg\,min}
\DeclareMathOperator{\asinh}{asinh}
\DeclareMathOperator{\Grad}{grad}
\DeclareMathOperator{\Law}{Law}
\DeclareMathOperator{\Dom}{Dom}

\DeclareMathOperator{\Prob}{\mathrm{Prob}}

\DeclareMathOperator{\Ker}{\mathrm{Ker}}

\DeclareMathOperator*{\sumsum}{\sum\!\sum}
\newcommand{\sumxly}{\sumsum_{(x,y)\in\X^2/2}}

\newcommand{\super}[1]{^{\scriptscriptstyle{(#1)}}}

\def\I{\mathcal{I}}

\def\A{\mathcal{A}}

\def\E{\mathcal{E}}
\def\H{\mathcal{H}}
\def\JJ{\mathbb{J}}
\def\KK{\mathbb{K}}
\def\L{\mathcal{L}}
\def\M{\mathcal{M}}
\def\MM{\mathbb{M}}
\def\P{\mathcal{P}}
\def\S{\mathcal{S}}
\def\T{\mathbb{T}}

\def\V{\mathcal{V}}
\def\W{\mathcal{W}}
\def\X{\mathcal{X}}

\def\Z{\mathcal{Z}}

\def\Q{\mathcal{Q}}
\def\RR{\mathbb{R}}
\def\TT{\mathbb{T}}
\def\NN{\mathbb{N}}
\def\ZZ{\mathbb{Z}}

\def\div{\mathop{\mathrm{div}}\nolimits} 
\def\Ddiv{\div}

\DeclareMathOperator{\dgrad}{ \overline{\scalebox{0.9}{\ensuremath\nabla}} }
\DeclareMathOperator{\ddiv}{\overline{\scalebox{0.9}{\textup{div}}}}

\date{\today}
\title{
Macroscopic Fluctuation Theory versus large-deviation-induced {GENERIC}
}
\author{D.R.\ Michiel Renger
\thanks{TU M{\"u}nchen, School of CIT, 85747 Garching, Germany. Email: \href{mailto:d.r.m.renger@tum.de}{d.r.m.renger@tum.de}} 
}

\begin{document}
\maketitle

\begin{abstract}
Recent developments in Macroscopic Fluctuation Theory show that many interacting particle systems behave macroscopically as a combination of a gradient flow with Hamiltonian dynamics. This observation leads to the natural question how these structures compare to the GENERIC framework. This paper serves as a brief survey of both fields and a comparison between them, including a number of example models to which the comparison results are applied.

\end{abstract}

\section{Introduction}
\label{sec:intro}

Whereas classic thermodynamics focuses on systems in equilibrium/detailed balance, the past decades have seen exciting new developments in non-equilibrium thermodynamics, studying systems that are driven out of detailed balance by boundary or bulk forces. 
We focus on two different frameworks to model and analyse the temporal evolution of such systems, namely: 
``General Equation for Non-Equilibrium Reversible-Irreversible Coupling'' (GENERIC) \cite{GrmelaOettinger1997I,GrmelaOettinger1997II,Ottinger12} and 
``Macroscopic Fluctuation Theory'' (MFT) \cite{BDSGJLL2002,BDSGJLL2015MFT,PattersonRengerSharma2024}.
The aim of this paper is to provide a (non-exhaustive) survey of both fields, forge them into the same level of description, and compare them with each other.

\subsection{Level of description}
\label{subsec:intro level}

In order to compare the two frameworks, we need to embed them in the same, abstract level of description, to account for the following differences. First, GENERIC is typically described in thermodynamic \emph{state} variables like particle density, momenta, temperature and internal energy, whereas MFT takes such states and particle \emph{fluxes} into account. Therefore, we will make the comparison at two different levels of description: the state and the (state and) flux formulation. At the state level, MFT rarely applies, since non-equilibrium fluxes are left out of the picture. At the flux level on the other hand, there are not many examples where GENERIC applies, although it can be expressed by considering cumulative fluxes as state variables, see \cite{Renger2018b} for such interpretation.


A second difference that needs to be taken into account is that MFT is a statistical theory, in the sense that all relevant objects are derived from large deviations of an underlying particle system. By contrast, GENERIC is originally proposed as a physical structure underlying evolution equations, although connections to large deviations have been studied, see~\cite{KLMP2020math}. We shall therefore compare the two frameworks, both in connection to a given (large-deviation) variational structure, and without such connection at the level of the evolution equation.

A third potential difference is that GENERIC combines gradient flows with a Hamiltonian system, whereas in MFT one typically has a gradient flow and a remaining ``antisymmetric'' flow. However, recent insights show that this antisymmetric flow is often Hamiltonian too~\cite{RengerSharma2023,PattersonRengerSharma2024}. Due to its similarity to GENERIC, we shall call such structure MANERIC (\underline{M}acroscopic-Fluctuation-Theory-based \underline{E}quation for \underline{N}on-\underline{E}quilibrium \underline{R}eversible-\underline{I}rreversible \underline{C}oupling). We are thus mainly interested in studying the similarities and differences between GENERIC and MANERIC. One such difference will be that not each system with a MANERIC structure will have a GENERIC structure and vice versa.

\subsection{Force vs. flux splitting}

The starting point of our analysis will be a (for now unspecified) microscopic system with random state or  mass density $\rho\super{n}(t)$ and random cumulative flux $W\super{n}(t)$, both related through some continuity equation $\dot\rho\super{n}(t)=d\phi_\rho \dot W\super{n}(t)$, where in practice the linear map $d\phi_\rho$ will often be the continuous or discrete divergence or a generalisation thereof. We assume that as $n\to\infty$ the noise vanishes and the process $(\rho\super{n}(t),W\super{n}(t))$ converges to the solution of the \emph{state-flux equations},
\begin{align}
  \dot \rho(t)=d\phi_{\rho(t)} j(t), &&  j(t):=\dot w(t) = j^0(\rho(t)).
\label{eq:limit evolution}
\end{align}
The first equation is essentially a continuity equation; the second one a constitutive law. We further assume a flux large-deviation principle (LDP, see~\eqref{eq:framework LDP}) with dynamic rate functional $\int_0^T\!\L(\rho(t),j(t))\,dt$, so that the minimiser of this non-negative action functional describes the typical behaviour~\eqref{eq:limit evolution} of the system. 

Corresponding to this action, we consider two different postulations that add physical meaning to the large deviations (all newly introduced symbols are described below):
\begin{align}
    \L(\rho,j)&=\Psi\big(\rho,j-\JJ(\rho)d\phi_\rho\tp d\E(\rho)\big)+\Psi^*\big(\rho,-\tfrac12d\phi_\rho\tp d\V(\rho)\big)+\langle \tfrac12 d\phi_\rho\tp d\V(\rho),j\rangle, \label{eq:L is GEN}\\    
    \L(\rho,j)&=\Psi\big(\rho,j\big)+\Psi^*\big(\rho,-\tfrac12d\phi_\rho\tp d\V(\rho) + F^\asym(\rho)\big)\notag\\
    &\hspace{16em}+\langle \tfrac12d\phi_\rho\tp d\V(\rho)-F^\asym(\rho),j\rangle \label{eq:L is MFT}.
\end{align}
In both formulations, $\Psi,\Psi^*$ are convex dual \emph{dissipation potentials} that generalise squared norms, see Definition~\ref{def:disspots}, and $\V$ is the quasipotential or free energy that drives the detailed balanced part of the dynamics. The adjoint $d\phi_\rho\tp$ of the continuity operator transforms forces $d\V(\rho),d\E(\rho)$ acting on velocities $\dot\rho$ to forces $\frac12d\phi_\rho\tp d\V(\rho), \frac12d\phi_\rho\tp d\E(\rho)$ acting on fluxes $j$.

In the flux \emph{GENERIC} postulation~\eqref{eq:L is GEN}, the non-detailed balanced dynamics is obtained by adding Hamiltonian dynamics $\JJ(\rho)d\phi_\rho\tp d\E(\rho)$ to the \emph{flux}, for some Poisson structure $\JJ(\rho)$ and conserving energy $\E$, in addition to other assumptions~\cite{Mielke2011GEN}. 
By contrast, in the MFT form~\eqref{eq:L is MFT}, the non-detailed balance dynamics is represented via an additional \emph{force} $F^\asym(\rho)$. This difference is also illustrated by considering the typical, zero-cost behaviour of \eqref{eq:L is GEN} and \eqref{eq:L is MFT}, respectively:
\begin{align}
   j^0(\rho)&=\overbrace{d_\zeta\Psi^*\big(\rho,-\tfrac12d\phi_\rho\tp d\V(\rho)\big)}^{=:j^{0\sym}(\rho)}+\overbrace{\JJ(\rho)d\phi_\rho\tp d\E(\rho)}^{=:j^{0\asym}(\rho)},
\notag\\
    j^0(\rho)&=d_\zeta\Psi^*\big(\rho,\underbrace{-\tfrac12d\phi_\rho\tp d\V(\rho)}_{=:F^\sym(\rho)} + F^\asym(\rho)\big).
        \label{eq:zero-cost MFT}
\end{align}
If $\Psi^*$ is a squared norm then the derivative $d_\zeta\Psi^*$ is a linear operator and the two formulations are equivalent. However, more general $\Psi^*$ leads to a \emph{nonlinear} response relations between forces and fluxes, and so one has to \emph{choose} whether to split the fluxes $j^0=j^{0\sym}+j^{0\asym}$ or the forces $F=F^\sym+F^\asym$. This choice has crucial consequences for the physical interpretation as well as the class of systems that can be cast in either form.

\subsection{Outline}

We end this introduction with presenting the abstract mathematical framework that will be used throughout the paper. Section~\ref{sec:MFT} contains a non-exhaustive overview of MFT, two new insights, and definitions including the newly introduced ``MANERIC''. Section~\ref{sec:GENERIC} is a non-exhaustive survey of GENERIC, in particular in connection with large deviations, and also contains the GENERIC definitions that we will need. In Section~\ref{sec:comparison} we compare the two frameworks at the large-deviation level, as well as on the level of the evolution equation. 

Finally we illustrate the theory by analysing 3 example models from an MFT and GENERIC perspective: the zero range process in Section~\ref{sec:zero range}, the Andersen Thermostat in Section~\ref{sec:Andersen} and a combination of zero-range with deterministic dynamics in Section~\ref{sec:zero+det}. The appendix contains a number of standard definitions and results that we include for completeness.

Table~\ref{tab:comparison table} provides a rough overview of the comparison between MANERIC and GENERIC.

\renewcommand{\arraystretch}{1.2}
\begin{table}[h!]
    \centering
    \begin{tabular}{p{0.45\linewidth} | p{0.45\linewidth}}
     \hline
     MANERIC  & (quasi-)GENERIC \\ 
     
     \hline\hline

     Splitting of forces, with & Splitting of fluxes, with \\
          
     generalised orthogonality of forces & non-interaction conditions \\
  
     \hline

     Hamiltonian energy conserved along full flow only after choosing the structure driven by total mass & Hamiltonian energy conserved along the full flow \\
     
     \hline

     Applies to flux cost $\L(\rho,j)$, & Applies to density cost $\hat\L(\rho,u)$,\\ 
     
     typically not to density cost $\hat\L(\rho,u)$  &  typically not to flux cost $\L(\rho,j)$ \\
     
     \hline
     
     \multicolumn{2}{l}{%
     $\L$ induces MANERIC      \hspace{5em}            $\impliedby$ $\L$ induces quasi-GENERIC }\\
     \multicolumn{2}{c}{(Corollary~\ref{cor:qGEN to MAN}) }\\
     \hline

     \multicolumn{2}{l}{%
       \begin{minipage}{0.45\textwidth}
         $j^0$ has MANERIC structure and Hamiltonian energy is total mass
       \end{minipage}%
                $\implies$  $u^0$ has quasi-GENERIC structure}\\
     \multicolumn{2}{c}{(Theorem~\ref{th:MAN to qGEN}) }\\
     \hline
     
     Applies to deterministic motion only if coupled to random motion in the same direction & Applies to deterministic motion only if coupled to detailed balance motion \\
     
     (Example Sections~\ref{sec:Andersen} and \ref{sec:zero+det})                       & (Proposition~\ref{prop:preGEN})\\
      
    \hline
    \end{tabular}
    \caption{A rough overview of the main similarities and differences between MANERIC and (quasi-)GENERIC.}
    \label{tab:comparison table}
\end{table}
\renewcommand{\arraystretch}{1}



\subsection{Mathematical framework}
\label{subsec:framework}

We use the abstract setting from~\cite{Renger2018b}, \cite{PattersonRengerSharma2024,AjjiChaoukiEsenGrmelaKlikaPavelka2023}. More precisely, we are given a differentiable Banach manifold $\Z$ of `states', another differentiable Banach manifold $\W$ of `fluxes', and a surjective differentiable operator $\phi:\W\to\Z$ mapping fluxes to states\footnote{In \cite{AjjiChaoukiEsenGrmelaKlikaPavelka2023}, $\Z$ is referred to as the base manifold, $T\W$ as the total manifold and $d\phi_\rho$ as the anchor map.}. The idea is that $\phi$ encodes a continuity equation $\rho=\phi\lbrack w\rbrack$ or at the level of tangents $\dot\rho=d\phi_w \dot w$, for example $\dot\rho=-\div\dot w$. In practice, the tangent space $T_w\W$ and differential $d\phi_w$ usually depends on $w$ through $\rho=\phi\lbrack w\rbrack$ only; hence by a slight abuse of notation we may write $T_\rho\W$ and $d\phi_\rho$ instead. We write $\langle \zeta,j\rangle$ for the dual pairing between \emph{fluxes} $j\in T_\rho\W$ and \emph{forces} $\zeta\in T_\rho^*\W$. 

Furthermore, we assume that all functions will depend on $w\in\W$ through $\rho:=\phi\lbrack w\rbrack$ only, so that we may write, again by a slight abuse of notation $(\rho,j)\in T\W$. This assumption can be interpreted as \emph{memorylessness}; for systems with rotationary motion for example, $w$ could store the winding number that does not play a role in the forces and energies. We refer to \cite{Renger2018b} for a more general framework that does allow for memory effects.

At the microscopic level, we study two random processes, $\rho\super{n}(t)$ on $\Z$ and $(\rho\super{n}(t),W\super{n}(t))$ on $\Z\times\W$ that are almost surely related through the continuity equation $\rho\super{n}(t)=\phi\lbrack W\super{n}(t)\rbrack$. Of course the processes $\rho\super{n}(t)$ and $(\rho\super{n}(t),W\super{n}(t))$ are basically the same, but the second one captures additional information in the flux variable. In practice all diffusion mobilities, jump rates and deterministic drifts depend on $\rho\in\Z$ only. This can be interpreted as microscopic memorylessness and implies macroscopic memorylessness.

To pass to the macroscopic level $n\to\infty$, we assume a \emph{large-deviation principle} (LDP) of the following form (see \cite{DemboZeitouni09,FengKurtz06} for the precise definition):
\begin{align}
  \Prob\Big( (\rho\super{n},W\super{n})\approx (\rho,w)\Big) &\stackrel{n\to\infty}{\sim} e^{-n\I_0(\rho(0))-n\int_0^T\!\L(\rho(t),\dot w(t))\,dt}, \qquad \dot\rho=d\phi_\rho \dot w,
\label{eq:framework LDP}
\end{align}
where $T>0$ is a fixed end time, $\L:T\W\to\lbrack0,\infty\rbrack$ is called the cost function, and $\I_0$ is the large-deviation rate of the random initial condition, which does not play a significant role. We consider $W\super{n}(t)$ to be a cumulative variable (e.g. the cumulative flux) and so we set the initial condition $W\super{n}(0)=0$ almost surely. This implies that the initial large deviations are actually $\I_0(\rho(0))$ if $w(0)=0$ and $\infty$ otherwise. The LDP~\eqref{eq:framework LDP} quantifies the unlikeliness to observe an atypical trajectory for large but finite $n$, and needs to be proven case-by-case. The assumption that the exponential takes the form of an action functional can be seen as a macroscopic version of Markovianity of the process $(\rho\super{n}(t),W\super{n}(t)$~\cite{FengKurtz06}. 

The main object of study is the LDP cost function $\L:T\W\to\lbrack0,\infty\rbrack$ from \eqref{eq:framework LDP} that measures the dynamic fluctuations. We assume that $\L$ satisfies the conditions for each $\rho\in\Z$,
\begin{enumerate}[(i)]
\item \emph{(microscopic memorylessness)} $\L$ depends on $w\in\W$ through $\rho:=\phi\lbrack w \rbrack$ only, so we can indeed write $\L=\L(\rho,j)$ for  $(\rho,j)\in T\W$, $\rho\in\Z$.
\item \emph{(convex and lsc)} $\L(\rho,\cdot)$ is its own convex bidual:
\begin{align}
    \H(\rho,\zeta):=\sup_{j\in T_\rho\W}\langle \zeta,j\rangle - \L(\rho,j),
    &&
    \L(\rho,j)=\sup_{\zeta\in T^*_\rho\W}\langle\zeta,j\rangle-\H(\rho,\zeta).
\label{eq:H dual L}
\end{align}
Systems for which $\L$ is not the convex dual of $\H$ (see for example \cite{JanevckaPavelka2018}) lie outside the scope of this paper.
\item \emph{(zero-cost flux)} $0=\min_{j\in T_\rho\W}\L(\rho,j):=\L(\rho,j^0(\rho))$ for some unique $j^0(\rho)$.
\end{enumerate}

As a consequence of the LDP~\eqref{eq:framework LDP}, the zero-cost flux encodes the limiting behaviour. More precisely, the random trajectories $(\rho\super{n}(t),W\super{n}(t))$ converge narrowly to the solution $(\rho(t),w(t))$ of the state-flux equations \eqref{eq:limit evolution} with initial conditions  $\rho(0)\in\argmin\I_0$ and $w(0)=0$.

Applying the ``contraction principle''~\cite[Thm.~4.2.1]{DemboZeitouni09} to \eqref{eq:framework LDP} also yields an LDP for the states:
\begin{equation*}
  \Prob\Big( \rho\super{n}\approx \rho\Big) \stackrel{n\to\infty}{\sim} e^{-n\I_0(\rho(0))-n\int_0^T\!\hat\L(\rho(t),\dot \rho(t))\,dt},
\end{equation*}
with cost function $\hat\L:T\Z\to\lbrack0,\infty\rbrack$ and its convex dual:
\begin{align}
  \hat\L(\rho,u):=\inf_{j\in T_\rho\W: d\phi_\rho j=u} \L(\rho,j), 
  &&
  \hat\H(\rho,\xi)=\H(\rho,d\phi_\rho\tp\xi). 
  \label{eq:hat L}                             
\end{align}
The two cost functions $\L$ and $\hat\L$ allow a comparison of MFT and GENERIC on the state manifold $\Z$ as well as on the flux manifold $\W$. 
Throughout the paper we work with a given triple $(\Z,\W,\phi)$ and a cost function $\L$ satisfying the conditions above. In principle, our abstract theory is purely macroscopic in the sense that the cost function $\L$ does not necessarily come from an LDP~\eqref{eq:framework LDP} corresponding to some random process.

\section{Survey on MFT}
\label{sec:MFT}

This section is meant as a concise, non-exhaustive overview of MFT and its more recent extensions to non-quadratic dissipation potentials. To keep the exposition snappy, we do not go too deeply into the history and literature. A great overview of the theory with quadratic dissipation can be found in the survey~\cite{BDSGJLL2015MFT}; for recent insights in the theory with non-quadratic potentials we refer to~\cite{PattersonRengerSharma2024}.

We shall work in the abstract setting from Section~\ref{subsec:framework}. Essential for MFT is the state-flux triple $(\Z,\W,\phi)$ instead of the state space $\Z$ only. On top of this structure we are given a sequence of Markov processes $(\rho\super{n}(t),W\super{n}(t))$ on $\Z\times\W$ with generator $\Q\super{n}:\Dom(\Q\super{n})\to C_b(\Z\times\W)$ so that $\rho\super{n}(t)$ is a Markov process on $\Z$, with generator $\hat\Q\super{n}:\Dom(\hat\Q\super{n})\to C_b(\Z)$. Furthermore, we assume the pathwise LDP~\eqref{eq:framework LDP} with given cost function $\L$.

The aim is to derive~\eqref{eq:L is MFT} in a meaningful way, where the symmetric and antisymmetric parts of the force will be determined through a time-reversal principle. Since this section is an overview, we simply refer to the original papers where the proofs can be found, and only present proofs for explicitly mentioned new insights.

\subsection{Micro to macro properties}

MFT is in essence a statistical theory, where macroscopic properties ($n\to\infty$) are derived from the microscopic system ($n\in\NN$). We therefore start with the following, almost trivial definitions for the microscopic Markov processes.
\begin{definition} For each $n\in\NN$,
\begin{enumerate}[(i)]
\item The measure $\Pi\super{n}\in\P(\Z)$ is called an \emph{invariant measure} of $\rho\super{n}(t)$ whenever $\int_{\P(\X)}\!(\hat\Q\super{n}G)(\rho)\,\Pi\super{n}(d\rho)=0$ for all test functions $G$ in the domain (or core) $\Dom(\hat\Q\super{n})\subset C_b(\Z)$.
\label{it:def inv meas}
\item Under initial condition $(\rho\super{n}(0),W\super{n}(0))\sim\Pi\super{n}\otimes\delta_0$ the Markov process
\begin{equation*}
  (\overleftarrow\rho\super{n}(t),\overleftarrow W\super{n})(t):=(\rho(T-t),W\super{n}(T)-W\super{n}(T-t))
\end{equation*}
is called the \emph{time-reversed} or adjoint  process (which has the same initial distribution).
\label{it:def time-reversed process}
\item We say that the process $(\rho\super{n}(t),W\super{n}(t))$ is in \emph{detailed balance} (with respect to $\Pi\super{n}$) if \footnote{We will not use the equivalent term ``reversibility'' to avoid confusion with the thermodynamic notion with the same name.}
\begin{equation}
  \Law(\overleftarrow\rho\super{n}(t),\overleftarrow W\super{n}(t))
  =
  \Law(\rho\super{n}(t),W\super{n}(t)).
\label{eq:def DB}
\end{equation}
\label{it:def DB}
\end{enumerate}
\label{def:MFT micro properties}
\end{definition}

\begin{remark}
Observe that Definition~\eqref{it:def inv meas} only applies to the state $\rho\super{n}(t)$; in practice, the cumulative flux $W\super{n}(t)$ does not have an invariant measure. Definitions~\eqref{it:def time-reversed process} and \eqref{it:def DB} are generalisations of the usual definitions to incorporate fluxes. Indeed $\overleftarrow\rho\super{n}(t)$ is the usual time-reversed process of $\rho\super{n}(t)$, and the flux variables can often be chosen so that detailed balance~\eqref{eq:def DB} is equivalent to the usual detailed balance of the process $\rho\super{n}(t)$, see for example~\cite[Sec.~4.1]{MaesNetocnyWynants2007Markov}, \cite[Sec. II.C]{BDSGJLL2015MFT}, \cite[Sec.~4.4]{Renger2018a}.
\end{remark}

The microscopic properties of Definition~\ref{def:MFT micro properties} can be translated to macroscopic ones.
\begin{proposition}[{\cite{BDSGJLL2015MFT},\cite[Sec.~4]{Renger2018a},\cite[Sec.~3]{PattersonRengerSharma2024}}] Let the following LDPs hold:
\begin{align*}
  \Prob\Big( (\rho\super{n},W\super{n})\approx (\rho,w)\Big) &\stackrel{n\to\infty}{\sim} e^{-n\I_0(\rho)-n\int_0^T\!\L(\rho(t),\dot w(t))\,dt}, \tag{\ref{eq:framework LDP}}\\
  \Prob\Big( (\overleftarrow\rho\super{n},\overleftarrow W\super{n})\approx (\rho,w)\Big) &\stackrel{n\to\infty}{\sim} e^{-n\I_0(\rho)-n\int_0^T\!\overleftarrow\L(\rho(t),\dot w(t))\,dt},\\
  \Pi\super{n}(\rho\super{n}\approx \rho) &\stackrel{n\to\infty}{\sim} e^{-n\V(\rho)}.
\end{align*}
Then:
\begin{enumerate}[(i)]
\item If for all $n\in\NN$, $\Pi\super{n}$ is an invariant measure for $\rho\super{n}$ then $\V$ satisfies the Hamilton-Jacobi equation (recall \eqref{eq:hat L}):
\begin{align}
  \hat\H\big(\rho,d\V(\rho)\big)=\H\big(\rho,d\phi\tp d\V(\rho)\big)=0,
  &&
  \inf_{\rho\in\Z}\V(\rho)=0
\label{eq:HJE}
\end{align}
for all $\rho\in\Z$ where $\V$ is differentiable.
\label{it:HJE}
\item The following time-reversal principle holds:
\begin{equation}
  \overleftarrow\L(\rho,j)=\L(\rho,-j)+\langle d\phi_\rho\tp d\V(\rho),j\rangle,
\label{eq:time-reversed L}
\end{equation}
for all $(\rho,j)\in T\W$ where $\V$ is differentiable.
\item If in addition the sequence of processes $(\rho\super{n}(t),W\super{n}(t))$ is in detailed with respect to $\Pi\super{n}$, then as a consequence of \eqref{eq:time-reversed L}, for all $(\rho,j)\in T\W$ where $\V$ is differentiable,
\begin{align}
  &\overleftarrow\L(\rho,j)=\L(\rho,j),  \notag\\
  &\L(\rho,j)=\L(\rho,-j)+\langle d\phi_\rho\tp d\V(\rho),j\rangle \quad{\text{and}}
  \label{eq:MFT time-reversal symmmetry}\\
  &d_j\L(\rho,0)=\tfrac12d\phi_\rho\tp d\V(\rho). \notag 
\end{align}
\label{it:MFT micro macro DB}
\end{enumerate}
\label{prop:MFT micro macro}
\end{proposition}
Note that we only require an invariant measure $\Pi\super{n}$ for $\rho\super{n}(t)$; in practice the cumulative fluxes $W\super{n}(t)$ do not have an invariant measure unless detailed balance holds.

From now on we shall always assume that $\V$ is a given quasipotential satisfying \eqref{eq:HJE}. The next decompositions do not require detailed balance, but are based on the time-reversal argument \eqref{eq:time-reversed L}.

\subsection{Force structure and decomposition}

The first step towards~\eqref{eq:L is MFT} is to decompose the cost $\L$ as a so-called force structure, following the terminology introduced in \cite{Renger2018a}.
\begin{definition} Let $\Psi:T\W\to\lbrack0,\infty\rbrack,\Psi^*:T^*\W\to\lbrack0,\rbrack$ be convex dual dissipation potentials, see Definition~\ref{def:disspots}, and $F:\Z\ni\rho\mapsto F(\rho)\in T_\rho^*\W$ a force field. We say that $\L$ \emph{induces the force structure} $(\Psi,F)$ whenever:
\begin{align}
  \L(\rho,j)=\Psi(\rho,j)+\Psi^*\big(\rho,F(\rho)\big)-\langle F(\rho),j\rangle, \label{eq:L Psi Psi*}
\end{align}
at all $j\in T_\rho\W$ and all $\rho$ for which the expressions are defined.
\label{def:force structure}
\end{definition}
Such decomposition generalises Onsager-Machlup/ $\Psi$-$\Psi^*$-formulations of a gradient flow, in the sense that the force does not need to be conservative.

\begin{proposition}[{\cite{BDSGJLL2015MFT},\cite{MielkePeletierRenger2014},\cite{Renger2018a},\cite{RengerZimmer2021},\cite{PattersonRengerSharma2024}}] At all states $\rho\in\Z$ where $\L(\rho,j)$ is differentiable in $j=0$, the cost $\L$ induces the unique force field $(\Psi,F)$, where
\begin{align*}
  F(\rho):=-d_j\L(\rho,0),&&
  \Psi^*(\rho,\zeta):=\H\big(\rho,\zeta-F(\rho)\big)-\H\big(\rho,-F(\rho)\big).
\end{align*}
Similarly at all states $\rho\in\Z$ where $\overleftarrow\L(\rho,j)$ is differentiable in $j=0$, the cost $\overleftarrow\L$ induces the unique force structure $(\overleftarrow\Psi,\overleftarrow F)$, and the time-reversed force structure $(\overleftarrow\Psi,\overleftarrow F)$ is related to the forward force structure $(\Psi,F)$ by, at all $\rho\in\Z$ for which the quasipotential $\V$ solving~\eqref{eq:HJE} is Gateaux differentiable,
\begin{align}
  \overleftarrow F(\rho):=-F(\rho)-d\phi_\rho\tp d\V(\rho),
  &&
  \overleftarrow \Psi(\rho,j)=\Psi(\rho,-j),
  &&
  \overleftarrow \Psi^*(\rho,\zeta)=\Psi^*(\rho,-\zeta),
\label{eq:reversed F}
\end{align}
\label{prop:force structure}
\end{proposition}
The existence and uniqueness of the force structure follows from a simple convexity argument, included in Prop.~\ref{prop:MPR} in the appendix. 
The relations~\eqref{eq:reversed F} follow as a consequence of the time-reversal principle~\eqref{eq:time-reversed L}, see the references above and \cite[Sec.~2.3]{PattersonRengerSharma2024} for the general argument. Interestingly, the dissipation potentials $\Psi,\Psi^*$ are the same for the forward and time-reversed dynamics -- apart from a minus sign, which becomes irrelevant if the dissipation potentials are even. In that case the difference between the forward and time-reversed dynamics is completely characterised by the forces $F,\overleftarrow F$.

\begin{remark}
Observe that the large-deviation cost $\overleftarrow\L$ of the time-reversed process need not be known a priori. Indeed, as soon as the quasipotential $\V$ satisfying the Hamilton-Jacobi equation~\eqref{eq:HJE} is known, $\overleftarrow{F}$ and $\overleftarrow\L$ can be reconstructed from \eqref{eq:reversed F}.
\end{remark}

\begin{remark}
As a consequence of the force structure, the term $\Psi^*(\rho,F(\rho))$, sometimes called the \emph{Fisher information}, characterises the long-time ergodic limit, see Subsection~\ref{subsec:applications}. 
\end{remark}

The next step towards~\eqref{eq:L is MFT} is to decompose the force into two components. Motivated by the previous result, those two components are:
\begin{align}
  F^\sym(\rho):=\tfrac12\big( F(\rho)+\overleftarrow F(\rho)\big)\stackrel{\eqref{eq:reversed F}}{=}-\tfrac12 d\phi_\rho\tp d\V(\rho),\;
  F^\asym(\rho):=\tfrac12\big( F(\rho)-\overleftarrow F(\rho)\big),
\label{eq:Fsym Fasym}
\end{align}
The symmetric (with respect to time reversal) part $F^\sym(\rho)$ of the force $F(\rho)$ is indeed related to detailed balance, and by \eqref{eq:reversed F}, this force is the negative derivative of the free energy $\V$. This derivative should be taken in the cumulative fluxes, which explains the appearance of the operator $d\phi_\rho\tp$:
\begin{align}
  \text{for } \check{\V}(w):=\V(\phi\lbrack w\rbrack), 
  && 
  d_w\check{\V}(w)=d\phi_\rho\tp d\V(\phi\lbrack  w\rbrack)=d\phi_\rho\tp d\V(\rho).
\label{eq:derivative wrt w}
\end{align}
The factor $1/2$ is crucial when working with non-quadratic dissipation potentials, and appears because the gradient itself quantifies the difference between the forward and backward dynamics~\eqref{eq:time-reversed L}, see for example~\cite{MaesRedigMoffaert2000}.

The antisymmetric part $F^\asym(\rho)$ is generally not of the form $d\phi_\rho\tp \hat F^\asym(\rho)$ for some vector field $\hat F^\asym(\rho)\in T_\rho^*\Z$. This again underlines the fact that one \emph{needs} to take fluxes into account to see the antisymmetric part of the force. For many models, $F^\asym(\rho)$ is independent of the state $\rho$. This is possibly related to the locality of the quasipotential and the absence of long-range correlations in the steady state, as shown in \cite[Sec.~V.C]{BDSGJLL2015MFT} for quadratic cost functions.

Summarising, under mild differentiability assumptions, \emph{any} system can be decomposed as in \eqref{eq:L is MFT}. Furthermore, in therms of trajectories, the zero-cost dynamics (minimising the action $\int_0^T\!\L(\rho(t),j(t))\,dt$) thus
satisfies the first of the following evolution equations:
\begin{align}
  \dot\rho(t)&=d\phi_{\rho(t)} d_\zeta\Psi^*\big(\rho(t),F^\sym(\rho(t))+F^\asym(\rho(t))\big),  &\text{(full dynamics)},
    \label{eq:MFT general dynamics}\\
  \dot\rho(t)&=d\phi_{\rho(t)} d_\zeta\Psi^*\big(\rho(t),F^\sym(\rho(t))\big),                                     &\text{(symmetric flow)},
     \label{eq:MFT symmetric flow}\\
  \dot\rho(t)&=d\phi_{\rho(t)} d_\zeta\Psi^*\big(\rho(t),F^\asym(\rho(t))\big),                                    &\hspace{-2cm}\text{(antisymmetric flow)}.
    \label{eq:MFT antisymm flow}
\end{align}
The second and third equations are interesting in their own right -- for example when symmetric/equilibrium and antisymmetric forces occur on different scales -- but also to understand the gap in FIR inequalities, see Corollary~\ref{cor:FIR} and different contributions to the free energy loss, see Corollary~\ref{cor:MFT 3 faces}. In particular, the second equation coincides with the detailed balance case and the resulting symmetric flow is essentially a gradient flow of the quasipotential $\V$, see Subsection~\ref{subsec:MFT symmetric flow}. The third equation is obtained by turning the symmetric (entropic) force $F^\sym$ off, and will often be Hamiltonian as we discuss in Subsection~\ref{subsec:MFT antisymmetric flow}. 

We stress again that one can \emph{not} simply add the symmetric and antisymmetric flows together to obtain the full flow (unless $\Psi^*$ is quadratic). In order to obtain the full flow, the forces rather than the fluxes or velocities need to added together. 

\begin{remark}
Clearly, the condition that $\L(\rho,j)$ is differentiable in $j=0$ is essential for the existence of the force $F(\rho)$ at that point. It is precisely this condition that typically breaks down for systems out of detailed balance if one does not include fluxes in the formulation, it also breaks down if one considers one-way fluxes instead of net fluxes, and it often breaks down if random dynamics is coupled to deterministic dynamics (as in Section~\ref{sec:Andersen}). The good news is that for most non-detailed-balanced systems in the net flux formulation, the differentiability of $\L(\rho,j)$ in $j=0$ is only problematic for points $\rho\in\Z$ that lie on the boundary of the admissible domain.
\label{rem:L diff in zero}
\end{remark}

\begin{remark} In case of a linear equation $\dot\rho(t)=A \rho(t)$ for some operator $A$, the distinction between $F^\sym$ and $F^\asym$ is very different from simply taking the ($\pi$-weighted) symmetric and skewsymmetric parts of $A$, as in \cite{HuetterSvendsen2013,DuongOttobre2023}. The two forces are more related to tilting techniques in large deviation theory, see for example~\cite[Sec.~3.1]{PattersonRengerSharma2024}.
\end{remark}

\subsection{Fisher information and ergodic limit (new insight)}
\label{subsec:ergodic}


Consider the force structure~\eqref{eq:L Psi Psi*}. For a specific model with quadratic convex dual dissipation potentials and $F^\asym=0$, it was found that the $\Psi(\rho,j)$-term characterises the \emph{short-time behaviour} of the system~\cite{Leonard2007}, whereas the work term $\langle F(\rho),j\rangle$ \cite{Leonard2007} is the next-order term of that short-time limit~\cite{AdamsDirrPeletierZimmer11,AdamsDirrPeletierZimmer13}. Based on a similar change of variables but for a different model, it was found that the Fisher information $\Psi^*(\rho,F(\rho))$ quantifies the convergence of the \emph{long-time} ergodic limit~\cite[Th.~6.1]{NueskenRenger2023}. Here we shall extend that proof to the general setting of our paper: with general convex dual dissipation potentials and in flux space. Although at this level of generality, the proof has to remain formal, as far as the author is aware, this is a new result. 

\begin{theorem} Let $(\rho\super{n}(t),W\super{n}(t))$ be a sequence of Markov processes in $\Z\times\W$ satisfying the continuity equation $\rho\super{n(t)}=\phi\lbrack W\super{n}(t)\rbrack$ almost surely. Assume that the sequence satisfies an LDP with rate functional $\I_0(\rho(0))+\int_0^T\!\L(\rho(t),\dot w(t))\,dt$, and that $\L$ allows the decomposition~\eqref{eq:L Psi Psi*}. Then the ergodic average:
\begin{equation*}
  \bar\rho\super{n}(T):=\frac1T\int_0^T\!\rho\super{n}(t)\,dt
\end{equation*}
satisfies an LDP as first $n\to\infty$ and then $T\to\infty$:
\begin{align}
  \Prob\big( \bar\rho\super{n}(T)\approx \bar\rho\big) \stackrel{\substack{n\to\infty\\T\to\infty}}{\sim} e^{-nT\Psi^*(\bar\rho,F(\bar\rho))}.
\label{eq:ergodic LDP}
\end{align}
\label{th:ergodic LDP}
\end{theorem}

\begin{proof}[Formal proof]
By the contraction principle~\cite[Thm.~4.2.1]{DemboZeitouni09}, $\bar\rho\super{n}(T)$ satisfies am LDP as $n\to\infty$ for $T>0$ fixed:
\begin{align}
  &\hspace{-1em}\Prob\big( \bar\rho\super{n}(T)\approx \bar\rho\big) \stackrel{n\to\infty}{\sim} e^{-n\bar\I_T(\bar\rho)},
  \label{eq:ergodic LDP fixed T}\\
  \bar\I_T(\bar\rho)&=\inf_{\substack{\rho:(0,T)\to\Z,\\T^{-1}\int_0^T\!\rho(t)\,dt=\bar\rho}}
    \int_0^T\!\hat\L(\rho(t),\dot\rho(t))\,dt\notag\\
  &\hspace{-1em}\stackrel{\eqref{eq:hat L},\eqref{eq:L Psi Psi*}}{=}
    \inf_{\substack{(\rho,w):(0,T)\to\Z\times\W,\\ \dot w\in T_\rho\W,\\ T^{-1}\int_0^T\!\rho(t)\,dt=\bar\rho}}
    \int_0^T\!\Big\{ \Psi(\rho(t),\dot w(t))+\Psi^*\big(\rho(t),F(\rho(t))\big)\notag\\[-1em]
  &\hspace{20em}  -\langle F(\rho(t)),\dot w(t)\rangle \Big\}\,dt\notag\\
  &=
    \inf_{\substack{(\rho,w):(0,1)\to\Z\times\W,\\ \dot w\in T_\rho\W,\\ \int_0^1\!\rho(t)\,dt=\bar\rho}}
    \Big\{
      \frac1T\int_0^1\!\Psi(\rho(t),\dot w(t))\,dt +T\int_0^1\!\Psi^*\big(\rho(t),F(\rho(t))\big)\,dt \notag\\[-1em]
      &\hspace{20em}  -\int_0^1\!\langle F(\rho(t)),\dot w(t)\rangle\,dt
    \Big\}.\notag
\end{align}
Therefore (in the appropriate topology),
\begin{equation*}
  \frac1T\bar\I_T(\bar\rho)\xrightarrow[T\to\infty]{\Gamma}
  \inf_{\substack{\rho:(0,1)\to\Z,\\ \int_0^1\!\rho(t)\,dt=\bar\rho}}
  \int_0^1\!\Psi^*\big(\rho(t),F(\rho(t))\big)\,dt
  =\Psi^*\big(\bar\rho,F(\bar\rho)\big),
\end{equation*}
which together with \eqref{eq:ergodic LDP fixed T} provides the precise formulation of \eqref{eq:ergodic LDP}.
\end{proof}

\begin{remark} Fluxes play an essential role in the proof, because without detailed balance and without fluxes, a decomposition of the type~\eqref{eq:L Psi Psi*} usually breaks down.
\label{rem:fluxes in MFT}
\end{remark}

\begin{remark} Since we are interested in the infinite time interval, it can be easily shown as in \cite[Th.~6.1]{NueskenRenger2023} that the initial large deviations $\I_0$ do not play a role, unless the model has degenerate states $\rho$ from which it can not `escape' in finite time with finite large-deviation cost. However, for systems with such degenerate states the pathwise LDP is also expected to fail, see \cite{AAPR2022}. For simplicity we have therefore omitted the functional $\I_0(\rho(0))$ from the proof above.
\end{remark}

\subsection{Symmetric flow}
\label{subsec:MFT symmetric flow}

We now study the behaviour of the symmetric flow \eqref{eq:MFT symmetric flow}, that is the case $F^\asym=0$. This is the classic and best understood case, and corresponds to a system in detailed balance\footnote{In principle, one might also have a microscopic system that is \emph{almost} in detailed balance, but the difference is negligible even on the large-deviation scale.} by Proposition~\ref{prop:MFT micro macro}\eqref{it:MFT micro macro DB}. We then have the following generalisation of the Onsager-Machlup principle~\cite{Onsager1953I} even far away from the steady state and in flux space.
\begin{theorem}[BDSGJLL2004GF,MielkePeletierRenger2014,Renger2018b]
If detailed balance holds (see Definition~\ref{def:MFT micro properties}\eqref{it:def DB}) then $F(\rho)=F^\sym(\rho)$ and $F^\asym(\rho)=0$ and so due to the fluctuation-dissipation relation~\eqref{eq:Fsym Fasym}, at all points $\rho\in\Z$ where $\V$ is differentiable and $\L(\rho,j)$ is differentiable in $j=0$, and all $j\in T_\rho\W$,
\begin{align}
  \L(\rho,j)&=\Psi(\rho,j)+\Psi^*\big(\rho,-\tfrac12 d\phi_\rho\tp d\V(\rho)\big) + \langle \tfrac12 d\phi_\rho\tp d\V(\rho),j\rangle,
\label{eq:L GF decomp}  
\intertext{and for the state cost function~\eqref{eq:hat L}, at all points $\rho\in\Z$ where $\V$ is differentiable and $\L(\rho,j)$ is differentiable in $j=0$, and all $u\in T_\rho\Z$,}
  \hat\L(\rho,u)&=\hat\Psi(\rho,u)+\hat\Psi^*\big(\rho,-\tfrac12 d\V(\rho)\big) + \langle \tfrac12 d\V(\rho),u \rangle,
\label{eq:hat L GF decomp}
\end{align}
where
\begin{align*}
  \hat\Psi(\rho,u):=\inf_{\substack{j\in T_\rho\W:\\u=d\phi_\rho j}} \Psi(\rho,j),
  &&\text{with convex dual}&&
  \hat\Psi^*(\rho,\xi)=\Psi(\rho,d\phi_\rho\tp \xi).
\end{align*}
\label{th:MPR}
\end{theorem}

The main point of this result is that the large-deviation cost $\L$ corresponding to microscopic fluctuations provides a unique and physically correct variational formulation of a \emph{gradient flow}. Indeed, the macroscopic zero-cost dynamics corresponding to \eqref{eq:hat L GF decomp} is
\footnote{If $\Psi^*$ is quadratic, then one can define the gradient as $\Grad_\rho\V(\rho):=d_\xi\hat\Psi^*(\rho, -\tfrac12 d\V(\rho)\big)$. Strictly speaking for non-quadratic $\Psi^*$, the equation should be called a `generalised gradient flow', or possibly a `derivative flow'.}
:
\begin{equation}
  \dot\rho(t)= d_\xi\hat\Psi^*\big(\rho(t), -\tfrac12 d\V(\rho(t))\big).
\label{eq:hat gradient flow}
\end{equation}
Similary, for \eqref{eq:L GF decomp}, the macroscopic zero-cost dynamics of the cumulative fluxes reads:
\begin{align*}
  \dot w(t)&= d_\zeta\Psi^*\big(\phi\lbrack w(t)\rbrack, -\tfrac12 d_w\check \V(w(t))\big),
 &&\text{with } \check{\V}(w)\text{ as in } \eqref{eq:derivative wrt w}.
\end{align*}

We emphasize that up until this point, all results (e.g. decomposition~\eqref{eq:MFT general dynamics}, the fluctuation-dissipation theorem~\eqref{eq:reversed F} and Theorem \ref{th:MPR}) hold in complete generality.

\begin{remark} In fact, most known gradient flows~\eqref{eq:hat gradient flow} have an underlying flux structure. For example, for Wasserstein gradient flows $\hat\Psi^*(\rho,\xi)=\tfrac12\lVert \xi\rVert^2_{H^1(\rho)}=\Psi^*(\rho,d\phi_\rho\tp \xi)$ for $\Psi^*(\rho,\zeta):=\tfrac12\lVert \zeta\rVert^2_{L^2(\rho)}$ and $\phi\lbrack w\rbrack:=\rho^0-\div w$ so that $d\phi_\rho\tp=\grad$. Similarly the (quadratic) dual dissipation potential for the ``discrete  Wasserstein gradient  flow''~\cite{Mielke2011GF,Maas2011} can be written as $\hat\Psi^*(\rho,\xi)=\tfrac12\dgrad\xi\cdot \KK(\rho)\dgrad\xi=:\Psi^*(\rho,\dgrad\xi)$ with the discrete divergence and gradient~\eqref{eq:discrete divergence} and Onsager matrix $\MM$ from \eqref{eq:zero-range MM}. Similarly, the (non-quadratic) dual dissipation potential for chemical reactions is of the form $\hat\Psi^*(\rho,\xi)=\Psi^*(\rho,\Gamma\tp\xi)$ where $\Gamma$ is the stoichiometric matrix~\cite{MPPR2017,MaasMielke2020}.
\label{rem:gradient flow flux form}
\end{remark}

\subsection{Antisymmetric flow}
\label{subsec:MFT antisymmetric flow}

We now study the antisymmetric flow~\eqref{eq:MFT antisymm flow}; this might be interpreted as the exact opposite of detailed balance. At least formally, by setting $F^\sym(\rho)=0$ we have completely turned off any entropic force, friction and/or dissipation, which violates the second law of thermodynamics on the long-time scale. Does this mean that we can expect periodic orbit solutions or even a Hamiltonian system (see Definition~\ref{def:Ham structure})?

By contrast to the generality of the force structure and symmetric flow, a general understanding of these ``antisymmetric flows'' is lacking. However, for many models, Hamiltonian structures for the antisymmetric flow have been shown case-by-case:
\begin{enumerate}
\item Independent particles on a fixed, finite graph, see~\cite[Sec.~4]{PattersonRengerSharma2024},
\item Zero-range processes on a finite graph (this paper Section~\ref{sec:zero range}),
\item Independent particles or zero-range processes on a finite graph as above, but with in- and outflow of particles at `boundary' nodes~\cite{RengerSharma2023},
\item Lattice gas models, under additional assumptions on the non-equilibrium drift~\cite[5.3]{PattersonRengerSharma2024}.
\end{enumerate}
It should be emphasized that the real challenge is to find a Hamiltonian structure \emph{without} extending the state space, otherwise any system can be written as a Hamiltonian system, see Appendix~\ref{app:any system Hamiltonian}.

The antisymmetric flow is interesting for different reasons. For example, it turns out that the antisymmetric flow precisely characterises the gap in the FIR inequality, see Corollary~\ref{cor:FIR} below. Morevoer, even though in practice entropic effects can never be fully shut down, apparently many systems can be seen as a combination of entropic and Hamiltonian dynamics. We shall formalise this concept in Subsection~\ref{subsec:MANERIC}.

\begin{remark} We restrict the analysis of the antisymmetric flow~\eqref{eq:MFT antisymm flow} to the state variables $\rho\in\Z$. The reason is that for non-equilibrium systems, the cumulative quantity $w(t)$ is expected to monotonically increase or decrease over time. This suggests that inertia is at play, and we can only expect to find a Hamiltonian system after extending the system with momenta or other variables.
\label{rem:cumflux inertia}
\end{remark}

\subsection{Further decomposition by generalised orthogonality}
\label{subsec:further decomp}

Plugging the decomposition of forces $F=F^\sym+F^\sym$ into \eqref{eq:L Psi Psi*} we arrive at \eqref{eq:L is MFT}, with corresponding zero-cost flux \eqref{eq:zero-cost MFT}. Thus, the contribution of the symmetric and asymmetric effects are still intertwined due to the Fisher information $\Psi^*\big(\rho,F^\sym(\rho)+F^\asym(\rho)\big)$. Recent developments show that this term itself can be decomposed into separate contributions using the notion of generalised orthogonality~\cite{KaiserJackZimmer2018,RengerZimmer2021,PattersonRengerSharma2024}. This result generalises the fact that in a Hilbert space $\frac12\lVert \zeta+ \tilde\zeta\rVert^2=\frac12\lVert \zeta\rVert^2+\langle\zeta,\tilde\zeta\rangle+ \frac12\lVert \tilde\zeta\rVert^2$ and the cross term drops out whenever $\zeta$ is orthogonal to $\tilde\zeta$. If $\Psi^*$ is not a squared Hilbert norm, a similar decomposition is still possible, but at the price modifying one of the two convex terms, and loosing bilinearity of the cross term:
\begin{align}
  \Psi^*\big(\rho,\zeta+\tilde\zeta\big) 
  &=\Psi^*_{\tilde\zeta} (\rho,\zeta) + \theta_\rho(\zeta,\tilde\zeta) + \Psi^*(\rho,\tilde\zeta) \label{eq:orthonal splitting}\\
  &=\Psi^*_{\zeta}(\rho,\tilde\zeta) + \theta_\rho(\tilde\zeta,\zeta) + \Psi^*(\rho,\zeta),\notag
\end{align}
where
\begin{align}
  \Psi^*_{\tilde\zeta}\big(\rho,\zeta\big)
    &:=\frac12\big\lbrack \Psi^*(\rho,\zeta+\tilde\zeta)+\Psi^*(\rho,\tilde\zeta-\zeta)\big\rbrack - \Psi^*(\rho,\tilde\zeta),\notag\\
  \theta_\rho(\zeta,\tilde\zeta)&:=\frac12\big\lbrack \Psi^*(\rho,\zeta+\tilde\zeta)-\Psi^*(\rho,\tilde\zeta-\zeta)\big\rbrack.
\label{eq:orth pairing}
\end{align}
Of course, in this formulation the decompositions are rather trivial, but the generalised orthogonality between the symmetric and antisymmetric force is less trivial.

\begin{proposition}[{\cite[Prop.~2.25]{PattersonRengerSharma2024}}] 
For all $\rho\in\Z$ where $\V$ is differentiable and $\L(\rho,j)$ is differentiable in $j=0$,
\begin{equation*}
  \theta_\rho\big(F^\sym(\rho),F^\asym(\rho)\big)=0,
\end{equation*}
and if $\Psi^*$ is even, that is $\Psi^*(\rho,-\zeta)=\Psi^*(\rho,\zeta)$, then also
\begin{equation*}
  \theta_\rho\big(F^\asym(\rho),F^\sym(\rho)\big)=0.
\end{equation*}
\label{prop:gen orth}
\end{proposition}


As a direct consequence of the force structure \eqref{eq:L Psi Psi*} and the generalised orthogonality (Proposition~\ref{prop:gen orth}) one obtains the following extremely powerful decomposition.
\begin{corollary}[{\cite[Th.~2.29 \& Cor.~2.31,]{PattersonRengerSharma2024}}]
For all $\rho\in\Z$ where $\V$ is differentiable and $\L(\rho,j)$ is differentiable in $j=0$, and all $j\in T_\rho\W$,
\begin{align}
  \L(\rho,j)
  &=\Psi(\rho,j)+\Psi^*\big(\rho,F^\asym(\rho)\big)-\langle F^\asym(\rho),j\rangle \notag\\
    &\hspace{9em} + \Psi^*_{F^\asym}\big(\rho,-\tfrac12d\phi_\rho\tp d\V(\rho)\big) + \langle \tfrac12d\phi_\rho\tp d\V(\rho),j\rangle, \label{eq:decomp1}\\
\intertext{and if $\Psi^*$ is even then also}
  \L(\rho,j)&=\Psi(\rho,j)+\Psi^*\big(\rho,-\tfrac12d\phi_\rho\tp d\V(\rho)\big)+ \big\langle \tfrac12d\phi_\rho\tp d\V(\rho),j\big\rangle \notag\\
    &\hspace{9em} + \Psi^*_{F^\sym}\big(\rho,F^\asym(\rho)\big) - \langle F^\asym(\rho),j\rangle.
    \label{eq:decomp2}
\end{align}
\label{cor:MFT decompositions}
\end{corollary}

Observe that the first three terms on the right-hand side of \eqref{eq:decomp2} constitute the force structure of a system in detailed balance; only the last two terms depend on $F^\asym$ and determine the non-equilibrium behaviour. Similarly, the first three terms of \eqref{eq:decomp1} constitute the force structure of a purely antisymmetric system, whereas the last two terms are due to entropic effects.

\begin{remark}
Decompositions \eqref{eq:decomp1} and \eqref{eq:decomp2} can be recast in the convex duality form:
\begin{align*}
  \L(\rho,j)
  &=\L_{F^\asym}(\rho,j) + \H_{F^\asym}\big(\rho,-\tfrac12d\phi_\rho\tp d\V(\rho)\big) + \langle \tfrac12d\phi_\rho\tp d\V(\rho),j\rangle \\ 
  &=\L_{F^\sym}(\rho,j) + \H_{F^\sym}\big(\rho,F^\asym(\rho)\big) - \langle F^\asym(\rho),j\rangle,
\end{align*}
where, motivated by force structures~\eqref{eq:L Psi Psi*} and the connection to large-deviation theory~\cite[Sec.~3]{PattersonRengerSharma2024}, the ``tilted'' cost functions are driven by a different force:
\begin{align}
  \L_G(\rho,j)&:=\Psi(\rho,j)+\Psi^*\big(\rho,G(\rho)\big)-\langle G(\rho),j\rangle, \notag\\
  \H_G(\rho,\zeta)&:=\sup_{j\in T_\rho\W}\langle\zeta,j\rangle-\L_G(\rho,j)=\Psi^*\big(\rho,\zeta+G(\rho)\big)-\Psi^*\big(\rho,G(\rho)\big).
\label{eq:modified H}
\end{align}
\end{remark}

\subsection{An interpretation of generalised orthogonality (new insight)}

The following result shows how the duality pairing $\theta_\rho$ can be used to drive a system out of detailed balance without changing the quasipotential. On a microscopic scale, this means that the tilting does not alter the invariant measure, see also~\cite[Sec.~5]{PattersonRengerSharma2024}.

\begin{proposition} Let $\L$ be given so that the corresponding force $F(\rho)=F^\sym(\rho)=-\tfrac12 d\phi_\rho\tp d\V(\rho)$ for some quasipotential $\V$ satisfying the Hamilton-Jacobi equation~\eqref{eq:HJE}, and let $G:\rho\mapsto G(\rho)\in T_\rho\W$ be another force field. Then $\V$ is also the quasipotential for the tilded cost $\L_{F+G}$ from~\eqref{eq:modified H} if and only if
\begin{equation*}
  \theta_\rho\big(F(\rho),G(\rho)\big)=0.
\end{equation*}
for all $\rho\in\Z$ where $\V$ is differentiable.
\label{prop:orthogonal same QP}
\end{proposition}
\begin{proof} First note that the duality pairing~\eqref{eq:orth pairing} is defined in terms of the dissipation potentials, and is thus independent of the force. Hence we can rewrite the dissipation potentials in terms of $\H_{F+G}$, defined by \eqref{eq:modified H},
\begin{align*}
  &2\theta_\rho\big(-\tfrac12 d\phi_\rho\tp d\V(\rho),G(\rho)\big)\\
  &\qquad\qquad=\Psi^*\big(\rho,-\tfrac12 d\phi_\rho\tp d\V(\rho)+G(\rho)\big)-\Psi^*\big(\rho,-\tfrac12 d\phi_\rho\tp d\V(\rho)-G(\rho)\big)\\
  &\qquad\qquad=\underbrace{\H_{F+G}\big(\rho,0\big)}_{=0} - \H_{F+G}\big(\rho,d\phi_\rho\tp d\V(\rho)\big).
\end{align*}
The orthogonality pairing is precisely $0$ when $\V$ satisfies the Hamilton-Jacobi equation for the system with force $F+G$.
\end{proof}

\subsection{Applications of the decomposition}
\label{subsec:applications}

Both decompositions~\eqref{eq:decomp1},\eqref{eq:decomp2} can be extremely helpful for different situations. We mention three applications.

\subsubsection*{Application 1: FIR inequality and gap}

Applying the Moreau-Fenchel-Young inequality to \eqref{eq:decomp1} and then minimising over all fluxes $j\in T_\rho\W$ satisfying the continuity equation $u=d\phi_\rho j$, we obtain the \emph{FIR inequality} \cite{DuongLamaczPeletierSharma17, HilderPeletierSharmaTse2020, RengerZimmer2021} on the flux cost and contracted cost~\eqref{eq:hat L},
\begin{corollary}[FIR inequality, {\cite[Cor.~2.34]{PattersonRengerSharma2024}}] For all $\rho\in\Z$ where $\V$ is differentiable and $\L(\rho,j)$ is differentiable in $j=0$, and for all $j\in T_\rho\W$,
\begin{align}
  \L(\rho,j) \geq \hat\L(\rho,d\phi_\rho j) \geq \Psi^*_{F^\asym}\big(\rho,-\tfrac12d\phi_\rho\tp d\V(\rho)\big) + \langle \tfrac12 d\phi_\rho\tp d\V(\rho),j\rangle.
\label{eq:FIR}
\end{align}
\label{cor:FIR}
\end{corollary}
\noindent These inequalities can be used to derive error estimates and compactness for proving singular limits and $\Gamma$-convergence \cite{DLPSS18,PeletierRenger2023}. 

Originally such inequalities were derived more directly. However, using MFT we obtain much more: a precise quantisation of the gap in flux space. Indeed in light of \eqref{eq:decomp1}, the FIR inequality \eqref{eq:FIR} becomes an equality precisely when
\begin{equation*}
  \Psi(\rho,j)+\Psi^*\big(\rho,F^\asym(\rho)\big)-\langle F^\asym(\rho),j\rangle=0,
\end{equation*}
that is, precisely if the dynamics follow the antisymmetric flow~\eqref{eq:MFT antisymm flow}.

For practical purposes, one would like to turn \eqref{eq:FIR} into a global inequality, i.e. on $\int_0^T\!\hat\L(\rho(t),\dot\rho(t))\,dt$ for arbitrary paths $\rho(t)$. This is usually done by an approximation argument that needs to be worked out case-by-case, see for example~\cite[Th.~1.6]{HilderPeletierSharmaTse2020}, \cite[Sec.~5]{RengerZimmer2021} and \cite[Part II.A]{Hoeksema23}.

\subsubsection*{Application 2: Three faces of the second law}
As another application of the decompositions~\eqref{eq:decomp1},\eqref{eq:decomp2}, we obtain \emph{explicit} expressions for the free energy loss $\langle \tfrac12d\phi_\rho\tp d\V(\rho),j\rangle$ and antisymmetric work $\langle F^\asym(\rho),j\rangle$ along the flow $j=j^0(\rho)$ (recall $\L(\rho,j^0(\rho))=0$ by assumption),
\begin{corollary} For all $\rho\in\Z$ so that $\L(\rho,j)$ is differentiable in $j=0$ and $\V$ is differentiable, and all $j\in T_\rho\W$,
\begin{align*}
  \big\langle \tfrac12 d\V(\rho),d\phi_\rho j^0(\rho)\big\rangle
  &=-\Psi\big(\rho,j^0(\rho)\big)-\Psi^*\big(\rho,F^\asym(\rho)\big)+\big\langle F^\asym(\rho),j^0(\rho)\big\rangle\\
  &\hspace{12em} - \Psi^*_{F^\asym}\big(\rho,-\tfrac12d\phi_\rho\tp d\V(\rho)\big) \leq0,\\
  - \big\langle F^\asym(\rho),j^0(\rho)\big\rangle
  &=-\Psi\big(\rho,j^0(\rho)\big)-\Psi^*\big(\rho,-\tfrac12d\phi_\rho\tp d\V(\rho)\big)-\langle \tfrac12d\phi_\rho\tp d\V(\rho),j^0(\rho)\rangle\\
  &\hspace{14em}- \Psi^*_{F^\sym}\big(\rho,F^\asym(\rho)\big) \leq0.
\end{align*}
\label{cor:MFT 3 faces}
\end{corollary}
\noindent The \emph{sign} of the two work expressions (and their sum) is known as the ``three faces of the second law of thermodynamics''~\cite{FreitasEsposito2022}. Again, the result above is much stronger since MFT allows to derive \emph{explicit expressions} rather than the negativity only.

\subsubsection*{Application 3: ergodic limit convergence speed (new insight)}

This application concerns numerical methods that are based on a combination of long-time and large-particle-number approximations~\cite{LiuWang2016Stein,NueskenRenger2023,CarilloChoiTotzeckTse2018,GarbunoNueskenReich2020}: 
\begin{equation*}
  \bar{\rho}_T\super{N} \stackrel{N,T\to\infty}{\approx} \pi.
\end{equation*}
We can now use Theorem~\ref{th:ergodic LDP} to quantise the convergence speed in $N,T$ of the probability. In practice one is rather interested in the convergence speed of the random variable $\bar\rho_T\super{N}$ itself, but with a predescribed, sufficiently high probability. This can be formally derived from Theorem~\ref{th:ergodic LDP} in the same fashion as central limit theorem are often derived from LDPs:
\begin{formalcorollary} For arbitrary $a>0$,
\begin{equation*}
  \Prob\Big( \Psi^*\big(\bar\rho\super{n}(T),F(\bar\rho\super{n}(T))\big) \geq \frac{a}{nT}\Big)\sim e^{-a}.
\end{equation*}
\end{formalcorollary}
\noindent So also in this setting the ``Fisher information'' $\Psi^*\big(\bar\rho,F(\bar\rho)\big)$ controls the convergence speed.

Let us now start with a system in detailed balance, i.e. $F(\rho)=F^\sym(\rho)=-1/2d\phi_\rho d\V(\rho)$, and ask the question how to break detailed balance to improve the convergence speed. Pick a force field $G(\rho)$ that is orthogonal to the original one, i.e. $\theta_\rho(F(\rho),G(\rho))=0$. Then by Proposition~\ref{prop:orthogonal same QP}, the quasipotential corresponding to the combined force $H=F+G$ is still $\V$, indicating that the invariant measure $\Pi\super{n}$ and the steady state $\pi$ that we are trying to approximate are unaltered. Moreover, 
\begin{align*}
 H^\sym=-\frac12d\phi_\rho d\V(\rho), &&\text{and}&& H^\asym(\rho)=G(\rho).
\end{align*}
By generalised orthogonality~\eqref{eq:orthonal splitting}, assuming $\Psi^*$ is even,
\begin{align*}
  \Psi^*(\rho,H(\rho))=\Psi^*\big(\rho,-\tfrac12d\phi_\rho\tp d\V(\rho)\big) + \Psi^*_{F(\rho)}(\rho,G(\rho)).
\end{align*}
The convergence speed of the altered system thus consists of two non-negative parts: the convergence speed we had for the original detailed balance model, and a bonus contribution due to the added detailed balance breaking force $G(\rho)$.

\subsection{MANERIC definitions}
\label{subsec:MANERIC}

So far, all results in this MFT section hold in high generality, apart from the discovery that for \emph{many} but not all systems the antisymmetric flow are Hamiltonian. This motivates the introduction of a new concept. 
Recall from Subsection~\ref{subsec:intro level} that the abbreviation MANERIC stands for a similar coupling between ``reversible'' and ``irreversible'' dynamics as in GENERIC, but now based on MFT, meaning that forces rather than fluxes are decomposed.

\subsubsection*{MANERIC induced by large deviations}

We start with the definition of MANERIC induced by a large-deviation cost function $\L$. Recall the notion of a Hamiltonian structure, see Definition~\ref{def:Ham structure} in the appendix.

\begin{definition}\phantom{a}
Let $\V$ be the quasipotential~\eqref{eq:HJE}, $(\Psi,F)$ be the force structure induced by $\L$, with symmetric and antisymmetric parts~\eqref{eq:Fsym Fasym}. We say that $\L$ \emph{induces the MANERIC structure} $(\Psi,F,\V,\hat\JJ,\E)$ whenever the antisymmetric flow~\eqref{eq:MFT antisymm flow} has the Hamiltonian structure $(\hat\JJ,\E)$ (at all points $\rho\in\Z$ for which $\V$ is differentiable and $\L(\rho,j)$ is differentiable in $j=0$).
\label{def:MANERIC}
\end{definition}
Four crucial observations need to be made:
\begin{itemize}
\item We do not require that the quasipotential $\V$ is Lyapunov along the full dynamics $\dot\rho(t)=d\phi_{\rho(t)} d_\zeta\Psi^*(\rho(t),F^\sym(\rho(t))+F^\asym(\rho(t)))$ as this holds \emph{automatically} by the three-faces-of-the-second-law Corollary~\ref{cor:MFT 3 faces}. 
\item By contrast, the energy $\E$ is only conserved along the antisymmetric flow $\dot\rho(t)=d\phi_{\rho(t)} d_\zeta\Psi^*(\rho(t),F^\asym(\rho(t)))=\hat\JJ(\rho) d\E(\rho(t)$, but generally not along the full dynamics. An additional GENERIC-type assumption $\Psi^*\big(\rho,\zeta+\theta d\phi_\rho\tp d\E(\rho)\big)\equiv  \Psi^*\big(\rho,\zeta\big)$ will not help in this matter, and is thus not included in our definition of MANERIC.

\item The definition above utilises a mixture of the fluxes $(\rho,j)\in T_\rho\W$ -- to identify the forces and dissipation potentials, and states $(\rho,u)\in T_\rho\Z$ -- to identify the Hamiltonian system. Metaphorically speaking, the concept of MANERIC lives in between the flux formulation and the state formulation. We shall see in the comparison Section~\ref{sec:comparison} that this interpretation is actually rather spot-on.

\item Since $(\hat\JJ,\E)$ only describes the antisymmetric flow after contracting to the state space, the full force $F$ can not be recovered from $\Psi,\V,\hat\JJ,\E$. Therefore we include it in the definition.
\end{itemize}

The definition of MANERIC applies to all four examples from Subsection~\ref{subsec:MFT antisymmetric flow}. This will be further investigated in Section~\ref{sec:zero range}.

\subsubsection*{MANERIC for an evolution equation}

We continue working in the geometric setup $\W,\Z,\phi$ of Subsection~\ref{subsec:framework} but now ignore any variational structure of a large-deviation cost function $\L$. Instead we focus purely on the state-flux equations~\eqref{eq:limit evolution}, for some given $j^0:\Z\to T_\rho\W$:
\begin{align}
  j(t)=j^0(\rho(t)), && \dot \rho(t)=d\phi_{\rho(t)} j(t).
\label{eq:state-flux equations2}
\end{align}
Of course, disregarding the variational structure gives more flexibility, and so more equations will actually have a MANERIC structure, defined as follows.

\begin{definition} We say that the state-flux equations \eqref{eq:state-flux equations2} \emph{have the MANERIC structure} $(\Psi,F,\V,\hat\JJ,\E)$ whenever there exists a cost function $\L:T\W\to\lbrack0,\infty\rbrack$ that induces the MANERIC structure $(\Psi,F,\V,\hat\JJ,\E)$ such and $\L$ is a variational formulation for $j^0$, that is $\L(\rho,j^0(\rho))=0$. 
\end{definition}

\begin{proposition} Let $\Psi,\Psi^*$ be convex dual dissipation potentials on $T\W$, $\JJ(\rho):T_\rho^*\Z\to T_\rho\Z$ be a Poisson structure, $\V,\E:\Z\to\RR$ be differentiable functionals with $\inf\V=0$. The state-flux equations~\eqref{eq:state-flux equations2} have the MANERIC structure $(\Psi,F,\V,\hat\JJ,\E)$ if and only if:
\begin{enumerate}[(i)]
\item $\Psi^*\big(\rho,d\phi_\rho\tp d\V(\rho)+d_j\Psi(\rho,j^0(\rho))\big) \equiv \Psi^*\big(\rho, d_j\Psi\big(\rho,j^0(\rho)\big)\big)$,
\item $d\phi_{\rho} d_\zeta\Psi^*\big(\rho, d_j\Psi(\rho,j^0(\rho))+\tfrac12 d\phi_\rho\tp d\V(\rho) \big) \equiv \hat\JJ(\rho)d\E(\rho)$.
\end{enumerate}
In that case, the corresponding cost function that induces the MANERIC structure $(\Psi,F,\V,\hat\JJ,\E)$ is uniquely given by
\begin{align*}
  \L(\rho,j)=\Psi(\rho,j)+\Psi^*\big(\rho, d_j\Psi(\rho,j^0(\rho)) \big) - \big\langle d_j\Psi(\rho,j^0(\rho)),j \big\rangle.
\end{align*}
\end{proposition}
\begin{proof}
``$\impliedby$''
Define
\begin{align*}
  F^\sym(\rho):=-\tfrac12 d\phi_\rho\tp d\V(\rho) \qquad\text{and}\qquad
  F^\asym(\rho):=d_j\Psi\big(\rho,j^0(\rho)\big)-F^\sym(\rho),\\
  \L(\rho,j):=\Psi(\rho,j)+\Psi^*\big(\rho,F^\sym(\rho)+F^\asym(\rho)\big)-\big\langle F^\sym(\rho)+F^\asym(\rho),j\big\rangle.
\end{align*}
Clearly this $\L$ induces the force structure $(\Psi,F^\sym+F^\asym)$. Moreover by assumption~\emph{(i)} and Proposition~\ref{prop:force structure},
\begin{multline*}
  \H\big(\rho,d\phi_\rho\tp d\V(\rho)\big)=\Psi^*\big(\rho,d\phi_\rho\tp d\V(\rho)+F^\sym(\rho)-F^\asym(\rho)\big)\\
                                          -\Psi^*\big(\rho,F^\sym(\rho)+F^\asym(\rho)\big)=0,
\end{multline*}
and so $\V$ is indeed the quasipotential and $F^\sym,F^\asym$ are indeed the symmetric and antisymmetric forces associated to $\L$. The second assumption says that the antisymmetric flow has the Hamiltonian structure $(\hat\JJ,\E)$.\\
``$\implies$''. The definitions of $F^\sym,F^\asym$ are now implications of the force structure and $\V$ being the quasipotential. The Hamilton-Jacobi equation~\eqref{eq:HJE} implies \emph{(i)}, and the antisymmetric flow being Hamiltonian implies \emph{(ii)}. The uniqueness of the cost function follows from the force structure $(\Psi,F)=(\Psi,d_j\Psi(\cdot,j^0(\cdot)))$.
\end{proof}

As for $\L$-induced MANERIC structures, $\V$ is again Lyapunov but $\E$ is generally not conserved along the full flow.

\section{Survey on large-deviation-induced GENERIC}
\label{sec:GENERIC}

We first introduce the definitions of GENERIC, pre-GENERIC and quasi-GENERIC for evolution equations and induced by large deviations. Next we derive relevant macroscopic properties from properties of the microscopic system, and finally discuss how pre-GENERIC and GENERIC are related to those properties. Most results in this section are taken from \cite{KLMP2020math}.

\subsection{GENERIC definitions}
\label{subsec:LDP GEN}

Following the chronology, we first introduce GENERIC structures for evolution equations before proceeding to GENERIC structures induced by large deviations. For consistency with the rest of the paper we work in the flux setting $(\Z,\W,\phi)$ of Subsection~\ref{subsec:framework}. We finally introduce the same structures in the state space $\Z$; all structures in flux and state space will be helpful in the comparison with MFT, and in particular the MANERIC structures.

\subsubsection*{GENERIC for an evolution equation}

The GENERIC framework is originally developed as a modelling tool to derive physically correct evolution equations. The idea goes back to \cite{Morrison1986metriplectic}, was reformulated in~\cite{GrmelaOettinger1997I,GrmelaOettinger1997II} and has been further developed, studied and applied since, see for example \cite{Ottinger2005,PavelkaKlikaGrmela2018} and the references therein. Translated to the flux setting, the definition is as follows.
\begin{definition}
Let $\Psi,\Psi^*$ be convex dual dissipation potentials on $T\W$ (Definition~\ref{def:disspots}), $\JJ(\rho):T_\rho^*\W\to T_\rho\W$ be a Poisson structure (Definition~\ref{def:Ham structure}), and $\V,\E:\Z\to\RR$ be differentiable functionals. For a given $j^0:\Z\ni\rho\mapsto T_\rho\W$, we say that the evolution equation $\dot w(t)=j^0\big(\phi\lbrack w(t)\rbrack\big)$ \emph{has the GENERIC structure}~$(\Psi,\V,\JJ,\E)$ whenever
\begin{enumerate}[(i)]
  \item $j^0(\rho) \equiv d_\zeta\Psi^*\big(\rho,-\tfrac12 d\phi_\rho\tp d\V(\rho)\big) + \JJ(\rho)d\phi_\rho\tp d\E(\rho)$,
  \item $\JJ(\rho)d\phi_\rho\tp d\V(\rho)\equiv0$,
    \label{it:eqGEN JV}
  \item $\Psi^*(\rho,\zeta+\theta d\phi_\rho\tp d\E(\rho))\equiv\Psi^*(\rho,\zeta)$ for all $\theta\in\RR$.
    \label{it:eqGEN PsiE}
\end{enumerate}
\label{def:equation GEN}
\end{definition}
The two ``non-interaction conditions'' \eqref{it:eqGEN JV}, \eqref{it:eqGEN PsiE} quarantee that along solutions, the free energy decays and the Hamiltonian energy is conserved. Indeed, setting $\rho(t)=\phi\lbrack w(t)\rbrack$,
\begin{align*}
  \frac{d}{dt} \V(\rho(t))&=\langle d\V(\rho(t)),\dot\rho(t)\rangle \\
   &=\big\langle d\phi_{\rho(t)}\tp d\V(\rho(t)),j^0(\rho(t))\big\rangle\\
   &=\overbrace{\big\langle d\phi_{\rho(t)}\tp d\V(\rho(t)),d_\zeta\Psi^*\big(\rho(t),-\tfrac12 d\phi_{\rho(t)}\tp d\V(\rho(t))\big)\big\rangle}^{\leq0} \\
   &\hspace{6em}+\underbrace{\big\langle d\phi_{\rho(t)}\tp d\V(\rho(t)),\JJ(\rho(t))d\phi_{\rho(t)}\tp d\E(\rho(t))\big\rangle}_{=0} \leq0
\end{align*}
due to the convexity of $\Psi^*$, and the skew-symmetry of $\JJ$ and non-interaction condition~\eqref{it:eqGEN JV}.
Similarly for the Hamiltonian energy,
\begin{align*}
  \frac{d}{dt} \E(\rho(t))
   &=\overbrace{\big\langle d\phi_{\rho(t)}\tp d\E(\rho(t)),d_\zeta\Psi^*\big(\rho(t),-\tfrac12 d\phi_{\rho(t)}\tp d\V(\rho(t))\big)\big\rangle}^{=0} \\
   &\hspace{6em}+\underbrace{\big\langle d\phi_{\rho(t)}\tp d\E(\rho(t)),\JJ(\rho(t))d\phi_{\rho(t)}\tp d\E(\rho(t))\big\rangle}_{=0} =0
\end{align*}
due to non-interaction condition~\eqref{it:eqGEN PsiE} and the skew-symmetry of $\JJ$.

\subsubsection*{GENERIC induced by large deviations}

For large-deviation-induced GENERIC, we are not only after properties of the zero-cost flow, but also after a connection to a given large-deviation cost function~$\L:T\W\to\lbrack0,\infty\rbrack$ of the form~\eqref{eq:L is GEN}, which is of course a much stronger statement. In the following definition we include two weaker definitions that will play a central role in this paper.

\begin{definition}\phantom{a}%
Let $\Psi,\Psi^*$ be convex dual dissipation potentials on $T\W$, $\JJ(\rho):T_\rho^*\W\to T_\rho\W$ be a Poisson structure, $\V,\E:\Z\to\RR$ be differentiable functionals and $b:\Z\ni\rho\mapsto b(\rho)\in T_\rho\W$ be a vector field.
\begin{itemize}
\item We say that $\L$ \emph{induces the GENERIC structure} $(\Psi,\V,\JJ,\E)$ whenever
\begin{enumerate}[(i)]
  \item $\L(\rho,j)\equiv\Psi\big(\rho,j-\JJ(\rho)d\phi_\rho\tp d\E(\rho)\big)+\Psi^*\big(\rho,-\tfrac12d\phi_\rho\tp d\V(\rho)\big) \\
         \indent\hspace{15em} + \big\langle \tfrac12d\phi_\rho\tp d\V(\rho),j-\JJ(\rho)d\phi_\rho\tp d\E(\rho)\big\rangle$,
  \item $\JJ(\rho)d\phi_\rho\tp d\V(\rho)\equiv0$, 
  \item $\Psi^*(\rho,\zeta+\theta d\phi_\rho\tp d\E(\rho))\equiv\Psi^*(\rho,\zeta)$ for all $\theta\in\RR$. 
\end{enumerate}

\item We say that $\L$ \emph{induces the quasi-GENERIC structure} $(\Psi,\V,\JJ,\E)$ whenever
\begin{enumerate}[(i)]
  \item $\L(\rho,j)\equiv\Psi\big(\rho,j-\JJ(\rho)d\phi_\rho\tp d\E(\rho)\big)+\Psi^*\big(\rho,-\tfrac12d\phi_\rho\tp d\V(\rho)\big) \\
         \indent\hspace{15em} + \big\langle \tfrac12d\phi_\rho\tp d\V(\rho),j-\JJ(\rho)d\phi_\rho\tp d\E(\rho)\big\rangle$,
  \item $\big\langle d\phi_\rho\tp d\V(\rho),j^0(\rho)\big\rangle\leq0$,
  \item $\Psi^*(\rho,\zeta+\theta d\phi_\rho\tp d\E(\rho))\equiv\Psi^*(\rho,\zeta)$ for all $\theta\in\RR$.
\end{enumerate}

\item We say that $\L$ \emph{induces the pre-GENERIC structure} $(\Psi,\V,b)$ whenever
\begin{enumerate}[(i)]
  \item $\L(\rho,j)\equiv\Psi(\rho,j-b(\rho))+\Psi^*\big(\rho,-\tfrac12d\phi_\rho\tp d\V(\rho)\big) \\
         \indent\hspace{15em} + \big\langle \tfrac12d\phi_\rho\tp d\V(\rho),j-b(\rho\big\rangle$,
  \item $\langle d\phi_\rho\tp d\V(\rho),b(\rho)\rangle\equiv0$.
\end{enumerate}
\end{itemize}

\label{def:GENs}
\end{definition}

In the definition of quasi-GENERIC, the non-interaction condition responsible for the decrease of the free energy $\V$ is simply replaced by the weaker condition that the free energy $\V$ decreases. In fact, in Section~\ref{subsec:comparison LDP level} we shall see that the weaker condition is often superfluous: if $\Psi,\Psi^*$ are even then $\V$ is also the MFT quasipotential, and so it will automatically decrease by Corollary~\ref{cor:MFT 3 faces}.

In the definition of pre-GENERIC, the Hamiltonian structure is replaced by a general drift. This structure does not have a strong physical interpretation like (quasi-) GENERIC, but it will act as an intermediate step towards derive large-deviation-induced GENERIC.

\subsubsection*{GENERIC in state space}

The structures on the state space $\Z$ are basically the same as the structures on flux space $\W$, but we include the definitions here for completeness and to fix the notation. Consistent with the rest of the paper, we notationally distinguish structures on state space from structures on flux space by a hat $\hat{}$ on top. For the structure induced by large deviations we are given a cost function $\hat\L:T\Z\to\lbrack0,\infty\rbrack$; in practice this will be connected to the cost on flux space through the contraction~\eqref{eq:hat L}. 

\begin{definition}
Let $\hat\Psi,\hat\Psi^*$ be convex dual dissipation potentials on $T\Z$, $\hat\JJ(\rho):T_\rho^*\Z\to T_\rho\Z$ be a Poisson structure, $\V,\E:\Z\to\RR$ be differentiable functionals and $\hat b:\Z\ni\rho\mapsto b(\rho)\in T_\rho\Z$ be a vector field.
\begin{itemize}

\item For a given $u^0:\Z\ni\rho\mapsto T_\rho\Z$, we say that the evolution equation $\dot \rho(t)=u^0(\rho(t))$ \emph{has the quasi-GENERIC structure}~$(\hat\Psi,\V,\hat\JJ,\E)$ whenever
\begin{enumerate}[(i)]
  \item $u^0(\rho) \equiv d_\xi\hat\Psi^*\big(\rho,-\tfrac12 d\V(\rho)\big) + \hat\JJ(\rho)d\E(\rho)$,
  \item $\big\langle d\V(\rho),u^0(\rho)\big\rangle\leq0$,
  \item $\hat\Psi^*(\rho,\xi+\theta d\E(\rho))\equiv\hat\Psi^*(\rho,\xi)$.
\end{enumerate}

\item We say that an $\hat\L:T\Z\to\lbrack0,\infty\rbrack$ \emph{induces the GENERIC structure} $(\hat\Psi,\V,\hat\JJ,\E)$ whenever
\begin{enumerate}[(i)]
  \item $\hat\L(\rho,u)\equiv \hat\Psi(\rho,u-\hat\JJ(\rho) d\E(\rho))+\hat\Psi^*\big(\rho,-\tfrac12 d\V(\rho)\big) \\
         \indent\hspace{18em}+ \big\langle \tfrac12 d\V(\rho),u-\hat\JJ(\rho)d\E(\rho)\big\rangle$,
  \item $\hat\JJ(\rho) d\V(\rho)\equiv0$, \eqnum\label{eq:ldpGEN JV}
  \item $\hat\Psi^*(\rho,\xi+\theta d\E(\rho))\equiv\hat\Psi^*(\rho,\xi)$. \eqnum\label{eq:ldpGEN PsiE}
\end{enumerate}

\item We say that an $\hat\L:T\Z\to\lbrack0,\infty\rbrack$ \emph{induces the quasi-GENERIC structure} $(\hat\Psi,\V,\hat\JJ,\E)$ whenever
\begin{enumerate}[(i)]
  \item $\hat\L(\rho,u)\equiv \hat\Psi(\rho,u-\hat\JJ(\rho) d\E(\rho))+\hat\Psi^*\big(\rho,-\tfrac12 d\V(\rho)\big) \\
         \indent\hspace{18em}+ \big\langle \tfrac12 d\V(\rho),u-\hat\JJ(\rho)d\E(\rho)\big\rangle$,
  \item $\big\langle d\V(\rho),u^0(\rho)\big\rangle\leq0$ (where $u^0(\rho)$ satisfies $\hat\L(\rho,u^0(\rho))=0$),
  \item $\hat\Psi^*(\rho,\xi+\theta d\E(\rho))\equiv\hat\Psi^*(\rho,\xi)$.
\end{enumerate}


\end{itemize}
\label{def:rho GEN}
\end{definition}

It generally holds that if a cost $\L$ on the flux space induces GENERIC, then the contracted cost $\hat\L$ on the state space also induces GENERIC. For more details we refer to \cite{Renger2018b}.

\subsection{Micro to macro properties}
\label{subsec:GEN micro to macro}

We are now ready to provide conditions under which a given cost $\L$ induces a (quasi-) GENERIC structure. Similar to Section~\ref{sec:MFT}, the first step is to derive suitable macroscopic properties of $\L$ from the microscopic dynamics. Recall the memorylessness of the generator $\Q\super{n}$ described in Subsection~\ref{subsec:framework}, Definition~\ref{def:MFT micro properties} of detailed balance and Definition~\ref{def:disspots} of convex dual dissipation potentials.

\begin{proposition}[\cite{KLMP2020math}]
Suppose that, for each $n\in\NN$, $\Q\super{n}=\Q\super{n}_\DB+\Q\super{n}_\drift$ so that:
\begin{enumerate}
\item $\Pi\super{n}$ is the invariant measure corresponding to $\Q\super{n}$ and $\Q\super{n}_\DB$,
\item The process with generator $\Q\super{n}_\DB$ is in detailed balance with respect to $\Pi\super{n}$,
\item The process with generator $\Q\super{n}_\drift$ is deterministic, i.e. 
\begin{equation*}
  (\Q\super{n}_\drift f)(\rho,w)=\langle d_w f(\rho,w),b(\rho)\rangle + \langle d_\rho f(\rho,w), d\phi_\rho b(\rho)\rangle
\end{equation*}
for some flux vector field $b(\rho)\in T_\rho\W$. 
\end{enumerate}
Let the following LDPs hold (here $(\rho\super{n}_\DB,W\super{n}_\DB)$ is the process with generator $\Q\super{n}_\DB$):
\begin{align*} 	
  \Prob\Big( (\rho\super{n},W\super{n})\approx (\rho,w)\Big) &\stackrel{n\to\infty}{\sim} e^{-n\I_0(\rho)-n\int_0^T\!\L(\rho(t),\dot w(t))\,dt}, \tag{\ref{eq:framework LDP}}\\
 \Prob\Big( (\rho\super{n}_\DB,W\super{n}_\DB)\approx (\rho,w)\Big) &\stackrel{n\to\infty}{\sim} e^{-n\I_0(\rho)-n\int_0^T\!\L_{\DB}(\rho(t),\dot w(t))\,dt},\\
 \Pi\super{n}(\rho\super{n}\approx \rho) &\stackrel{n\to\infty}{\sim} e^{-n\V(\rho)}.
\end{align*}
Then for all $(\rho,j)\in T\W$ where $\V$ is differentiable,
\begin{align*}
  &\L_\DB(\rho,j)=\L\big(\rho,j+b(\rho)\big), \\
  &\L\big(\rho,j+b(\rho)\big)=\L\big(\rho,-j+b(\rho)\big) + \langle d\phi_\rho\tp d\V(\rho),j\rangle,\\
  &\langle d\phi_\rho\tp d\V(\rho),b(\rho)\rangle=0, \quad\text{and}\\
  &d_j\L\big(\rho,b(\rho)\big)=\tfrac12d\phi_\rho\tp d\V(\rho),\\ 
\end{align*}
\label{prop:GEN micro to macro}
\end{proposition}
\begin{proof} The proof is taken from \cite{KLMP2020math}, adapted to the flux setting, and shortened by explicitly using the Hamilton-Jacobi equation~\eqref{eq:HJE}. Since the process with generator $Q\super{n}_\drift$ is deterministic, it also satisfies an LDP similar to \eqref{eq:framework LDP}, with cost function $\L_\drift(\rho,j)=0$ for $j=b(\rho)$ and $\L_\drift(\rho,j)=\infty$ for $j\neq b(\rho)$, and convex dual $\H_\drift(\rho,\zeta)=\langle \zeta,b(\rho)\rangle$. Since the generators are added, we also have $\H=\H_\DB+\H_\drift$, where $\H_\DB$ is the convex dual to $\L_\DB$; taking the convex dual yields the first claim $\L(\rho,j)=\L_\DB(\rho,j-b(\rho))$. 

The second claim follows from applying \eqref{eq:MFT time-reversal symmmetry} to $\L_\DB$, and the last claim is found by differentiating the second in $j=0$.

Finally, to show the orthogonality between $b(\rho)$ and $d\phi_\rho\tp d\V(\rho)$, note that $\Pi\super{n}$ is automatically also the invariant measure corresponding to $Q\super{n}_\drift$, and so by Proposition~\ref{prop:MFT micro macro}\eqref{it:HJE} the Hamilton-Jacobi equation $\H_\drift(\rho,d\phi_\rho\tp d\V(\rho))=0$ holds.
\end{proof}

\subsection{From large deviations to pre-GENERIC}

The outcome of Proposition~\ref{prop:GEN micro to macro} allows us to derive pre-GENERIC induced by a cost $\L$.

\begin{proposition}[\cite{KLMP2020math}] Let $\V$ be a differentiable functional and $b:\Z\ni\rho\mapsto b(\rho)\in T_\rho\W$ be a flux vector field. The following statements are equivalent at all points $\rho\in\Z$ where the expressions are well-defined:
\begin{itemize}
\item $\L\big(\rho,j+b(\rho)\big)\equiv\L\big(\rho,-j+b(\rho)\big) + \langle d\phi_\rho\tp d\V(\rho),j\rangle$ and $\langle d\phi_\rho\tp d\V(\rho),b(\rho)\rangle\equiv0$,
\item $\L$ induces the pre-GENERIC structure $(\Psi,\V,b)$ for some unique convex dual dissipation potentials $\Psi,\Psi^*$ on $T\W$ and the potentials are even, that is,
\begin{align*}
  &\Psi(\rho,-j)\equiv\Psi(\rho,j), \qquad\qquad \Psi^*(\rho,-\zeta)\equiv\Psi^*(\rho,\zeta).
\end{align*}
\end{itemize}
In that case the dual potential is explicitly given by:
\begin{equation*}
  \Psi^*(\rho,\zeta):=\H\big(\rho,\zeta+\tfrac12 d\phi_\rho\tp d\V(\rho)\big)-\H\big(\rho,\tfrac12 d\phi_\rho\tp d\V(\rho) -\langle\zeta,b(\rho)\rangle.
\end{equation*}
\label{prop:preGEN}
\end{proposition}
\begin{proof}
Again, we provide a short proof for completeness, see also \cite{KLMP2020math}.\\
``$\impliedby$'': the first condition is easily checked, and the orthogonality relation follows from Proposition~\ref{prop:incomplete L}. \\
``$\implies$'': define $\L_\DB(\rho,j):=\L\big(\rho,j+b(\rho)\big)$ so that $\L_\DB(\rho,j)=\L_\DB(\rho,-j)+\langle d\phi_\rho\tp d\V(\rho)\rangle$. The result then follows from Proposition~\ref{prop:MPR}.
\end{proof}

\subsection{Large deviation-induced pre-GENERIC to GENERIC}

If $\L$ indeed induces a pre-GENERIC structure, can we conclude that $\L$ actually induces a (quasi-) GENERIC structure? With a lot of luck, the drift $b(\rho)$ may already be of the Hamiltonian form $\JJ(\rho)d\E(\rho)$, so that by definition $\L$ will induce a (quasi-) GENERIC structure. Unfortunately, there are not many examples where this actually holds. As explained in Remark~\ref{rem:cumflux inertia}, in the non-detailed balance setting, one expects inertia, in which case the dynamics can only be a Hamiltonian system, either 1) after contracting to the state space $\Z$ or 2) after extending the space with an additional variable like energy or momentum (see Appendix~\ref{app:any system Hamiltonian}). 


\subsubsection*{Option 1: pre-GENERIC to (quasi-) GENERIC in state space}

The following is rather trivial and follows by definition, but we include it to stress its relevance. Recall that that $\hat\L:T\Z\to\lbrack0,\infty\rbrack$ denotes the contracted cost~\eqref{eq:hat L}.

\begin{corollary} If $\hat\L$ induces the pre-GENERIC structure $(\hat\Psi,\V,\hat b)$ and $\hat b(\rho)=\hat\JJ(\rho) d\E(\rho)$ for some Poisson structure $\hat\JJ(\rho):T_\rho^*\Z\to T_\rho\Z$ and differentiable energy $\E:\Z\to\RR$, and the non-interaction condition \eqref{eq:ldpGEN PsiE} holds, then $\hat\L$ induces the quasi-GENERIC structure $(\hat\Psi,\V,\hat\JJ,\E)$. \\ If in addition the non-interaction condition \eqref{eq:ldpGEN JV} holds then $\hat\L$ induces the GENERIC structure $(\hat\Psi,\V,\hat\JJ,\E)$.
\label{cor:pGEN to GEN state}
\end{corollary}

The results from Subsection~\ref{subsec:GEN micro to macro} show that these structures typically hold whenever the microscopic dynamics consists of a detailed balance part coupled to a deterministic Hamiltonian system.

Following~\cite{KLMP2020math} we stress that if indeed $\hat b(\rho)=\hat\JJ(\rho) d\E(\rho)$ for some $\JJ,\E$, it is not known whether these correspond to the variational structure of $\hat\L$ in any sensible way, and therefore such $\hat\JJ,\E$ will not be unique.

\subsubsection*{Option 2: pre-GENERIC to GENERIC in an extended state space}

In full generality, it turns out that pre-GENERIC is sufficient for $\L$ to induce GENERIC, but only \emph{after} extendsing the space with additional variables. To emphasize the extension, we write all extended expressions with a $\sim$ above them. 

\begin{proposition}[\cite{KLMP2020math}]
Assume that $\L$ induces the pre-GENERIC structure $(\Psi,\V,b)$. Define the extended state-flux triple and cost as ($f$ is a placeholder for $\dot e(t)$),
 \begin{align*}
  \tilde\Z:=Z\times\RR, && \tilde\W:=\W\times\RR, &&\tilde\phi\lbrack w,e\rbrack:=(\phi\lbrack w\rbrack, e),
\end{align*}
and let, for some $f^0:\tilde\W\to\RR$,
\vspace{-1em}
\begin{equation*}
  \tilde\L(\tilde\rho,\tilde\jmath)=\tilde\L\big((\rho,e),(j,f)\big):=
    \begin{cases} 
      \L(\rho,j), &f=f^0(\rho,e),\\
      \infty,     &f\neq f^0(\rho,e).
    \end{cases}
\end{equation*}
Then there exists a $f^0$ so that $\tilde\L$ induces a GENERIC structure $(\tilde\Psi,\tilde\V,\tilde\JJ,\tilde\E)$.
\label{prop:pGEN to GEN}
\end{proposition}
The paper \cite{KLMP2020math} provides a few alternative constructions. For the first construction, $f^0(e)=0$, meaning that we add a trivial energy variable $e(t)$ that by definition is not allowed to vary. The construction then builds upon Proposition~\ref{prop:Ham struct 1}. For the second construction, one assumes that the drift $b(\rho)$ is already of Hamiltonian type $\JJ(\rho)d\E(\rho)$, but the non-interaction condition is violated, so that $\E$ is not conserved along the full dynamics. In that case one adds the missing energy (due to exchange with an external heat bath) back into the system, and redefines the free energy $\V$ and Hamiltonian energy $\E$ in a smart way; this construction is detailed in \cite{DuongPeletierZimmer13} and \cite[Sec.~4.2]{KLMP2020math}. It is not clear to the author whether a similar construction can be done using the construction of Proposition~\ref{prop:Ham struct 2}.

Irregardless of the choice of $f^0$, one obtains that the GENERIC structure $(\tilde\Psi,\tilde\V,\tilde\JJ,\tilde\E)$ is induced by $\tilde\L$ in the sense that
\begin{align*}
  \tilde\L(\tilde\rho,\tilde\jmath)&=\tilde\Psi\big(\tilde\rho,\tilde\jmath-\tilde\JJ(\tilde\rho)d\tilde\phi_{\tilde\rho}\tp d\tilde\E(\tilde\rho)\big)+\tilde\Psi^*\big(\tilde\rho,-\tfrac12d\tilde\phi_{\tilde\rho}\tp d\tilde\V(\tilde\rho)\big)+\langle \tfrac12 d\tilde\phi_{\tilde\rho}\tp d\tilde\V(\tilde\rho),\tilde\jmath\rangle,
\end{align*}
which is \emph{not} the same as saying that $\L$ induces a GENERIC structure, as in Corollary~\ref{cor:pGEN to GEN state} and \eqref{eq:L is GEN} in the introduction. In fact, the Hamiltonian system $\tilde\JJ,\tilde\E$ is not very meaningful in itself, since \emph{any} evolution has a Hamiltonian structure after such extension (see Appendix~\ref{app:any system Hamiltonian}.

\section{Comparing GENERIC and MANERIC}
\label{sec:comparison}

In Sections~\ref{sec:MFT} and \ref{sec:GENERIC} we discussed and introduced the notions of large-deviation induced GENERIC and (MFT-based) MANERIC respectively. We are now ready to compare the two notions. In the first subsection we focus on what we can deduce from a given cost $\L:T\W\to\lbrack0,\infty\rbrack$, and we shall see that quasi-GENERIC implies MANERIC. 
In the second subsection we focus on the state-flux equations, and will see that MANERIC implies quasi-GENERIC if one allows to alter the variational structure.

\subsection{LDP level: quasi-GENERIC to MANERIC}
\label{subsec:comparison LDP level}

The following result shows that the free energy of a (pre-)GENERIC system generally coincides with the free energy or quasipotential as used in MFT. But although both frameworks are based on dissipation potentials, these turn out to be different, so we will need to distinguish them notationally.

\begin{lemma} If $\L$ induces a pre-GENERIC structure $(\Psi_\GEN,\V,b)$ with $\inf\V=0$ then $\L$ induces the force structure $(\Psi_\MFT,F)$ (see Definition~\ref{def:force structure}) with, at all points where the derivatives are well-defined:
\begin{equation}
  F(\rho)=-\tfrac12d\phi_\rho\tp d\V(\rho)-d_j\Psi_\GEN(\rho,-b(\rho)),
\label{eq:pGEN force}
\end{equation}
\vspace{-2em}
\begin{align*}
 \Psi_\MFT^*(\rho,\zeta)&=\Psi_\GEN^*\big(\rho,\zeta+d_j\Psi_\GEN(-b(\rho))\big)-\Psi_\GEN^*\big(\rho,d_j\Psi_\GEN(-b(\rho))\big)\\
  &\hspace{21em} +\langle \zeta,b(\rho)\rangle,\\
  \Psi_\MFT(\rho,j)
  &=\Psi_\GEN\big(\rho,j-b(\rho)\big)-\Psi_\GEN\big(\rho,-b(\rho)\big) - \big\langle d_j\Psi_\GEN(\rho,-b(\rho)),j\big\rangle.
\end{align*}
If in addition $\Psi_\GEN$ is even, that is $\Psi_\GEN(\rho,j)\equiv\Psi_\GEN(\rho,-j)$, then
\begin{enumerate}[(i)]
\item $\V$ is a quasipotential associated to $\L$,
\item The symmetric and antisymmetric components of the force are
\begin{align}
  F^\sym(\rho)&=-\tfrac12d\phi_\rho\tp d\V(\rho), \notag\\
  F^\asym(\rho)&=-d_j\Psi_\GEN(\rho,-b(\rho))=d_j\Psi_\GEN(\rho,b(\rho)),
\label{eq:forces from pGEN}
\end{align}
\item The dissipation potentials can be rewritten as
\begin{align}
 \Psi_\MFT^*(\rho,\zeta)&=\Psi_\GEN^*\big(\rho,\zeta-F^\asym(\rho)\big)-\Psi_\GEN^*\big(\rho,-F^\asym(\rho))\big) +\langle \zeta,b(\rho)\rangle, \notag\\
  \Psi_\MFT(\rho,j)
  &=\Psi_\GEN\big(\rho,j-b(\rho)\big)-\Psi_\GEN\big(\rho,-b(\rho)\big) + \big\langle F^\asym(\rho),j\big\rangle.
\label{eq:PsiMFT from PsiGEN}
\end{align}
\end{enumerate}
Moreover, if $\Psi_\GEN$ is quadratic, i.e. $\Psi_\GEN(\rho,j)=\frac12\langle G(\rho)j,j\rangle$ for some positive semidefinite operator $G(\rho):T_\rho\W\to T_\rho^*\W$, then $\Psi_\GEN=\Psi_\MFT$.
\label{lem:pGEN to MAN}
\end{lemma}

\begin{proof} By the pre-GENERIC assumption, the convex dual of $\L$ is 
\begin{equation*}
  \H(\rho,\zeta)=\Psi_\GEN^*\big(\rho,\zeta-\tfrac12d\phi_\rho\tp d\V(\rho)\big) - \Psi_\GEN^*\big(\rho,-\tfrac12d\phi_\rho\tp d\V(\rho)\big) + \langle\zeta,b(\rho)\rangle.
\end{equation*}
The expressions for $F$ and $\Psi^*_\MFT$ then follow from the formulas in Proposition~\ref{prop:force structure}. Taking the convex dual of $\Psi_\MFT^*$ yields
\begin{multline*}
  \Psi_\MFT(\rho,j)=\Psi_\GEN\big(\rho,j-b(\rho)\big)+\Psi_\GEN^*\big(\rho,d_j\Psi_\GEN(\rho,-b(\rho))\big) \\- \langle d_j\Psi_\GEN(\rho,-b(\rho)),j-b(\rho)\rangle.
\end{multline*}
From this we obtain the claimed expression for $\Psi_\MFT(\rho,j)$ by explicitly calculating the maximiser in the convex dual:
\begin{align*}
  \Psi_\GEN(\rho,-b(\rho))&=\sup_{\zeta\in T_\rho^*\W}\langle \zeta,-b(\rho)\rangle-\Psi^*_\GEN(\rho,\zeta)\\
  &=\big\langle d_j\Psi_\GEN(\rho,-b(\rho)),-b(\rho)\big\rangle-\Psi^*_\GEN\big(\rho,d_j\Psi_\GEN(\rho,-b(\rho))\big).
\end{align*}

Now assume that $\Psi_\GEN$ is even, implying that $\Psi_\GEN^*$ is even. Together with the non-interaction condition from the definition of pre-GENERIC,
\begin{align*}
  &\H\big(\rho,d\phi_\rho\tp d\V(\rho)\big)\\
  &\quad	=\Psi_\GEN^*\big(\rho,\tfrac12d\phi_\rho\tp d\V(\rho)\big) - \Psi_\GEN^*\big(\rho,-\tfrac12d\phi_\rho\tp d\V(\rho)\big) +\big\langle d\phi_\rho\tp d\V(\rho),b(\rho))\big\rangle=0,
\end{align*}
and so $\V$ is indeed a quasipotential. The explicit expressions for $F^\sym,F^\asym$ and $\Psi_\MFT^*,\Psi_\MFT^*$ follow as a consequence. Assuming that $\Psi_\GEN$ is quadratic, it is easily seen that indeed $\Psi_\MFT=\Psi_\GEN$.
\end{proof}
The converse of the last statement is not true: one can construct rather pathological examples where $b,F^\asym\neq0$ and $\Psi_\GEN=\Psi_\MFT$ but $\Psi_\GEN$ is not quadratic.

Quadratic dissipation potentials are relatively restrictive in the sense that they arise from microscopic processes that are approximately driven by white noise in continuous space. Even dissipation potentials are much more abundant, although non-even dissipation potentials may appear occasionally~\cite[Remark 2.13]{PattersonRengerSharma2024}.

The previous lemma shows that with even dissipation the quasipotential of a quasi-GENERIC structure is known so that we can connect quasi-GENERIC to MFT. This leads to the following result.

\begin{corollary} If $\L$ induces a quasi-GENERIC structure $(\Psi_\GEN,\V,\JJ,\E)$ with even $\Psi_\GEN,\Psi_\GEN^*$ and $\inf\V=0$, then $\L$ induces the MANERIC structure $(\Psi_\MFT,F,\V,\hat\JJ,\E)$ at all points where $\Psi_\MFT,\Psi^*_\MFT$ given by \eqref{eq:PsiMFT from PsiGEN} are well-defined, $F$ is given by \eqref{eq:pGEN force} and
\begin{equation}
  \hat\JJ(\rho):= d\phi_\rho \JJ(\rho)d\phi_\rho\tp.
\label{eq:contracted JJ}
\end{equation}
In addition, the following non-interaction condition also holds:
\begin{equation*}
  \Psi^*_\MFT\big(\rho,\zeta+\theta d\phi_\rho\tp d\E(\rho)\big)\equiv  \Psi^*_\MFT\big(\rho,\zeta\big).
\end{equation*}
\label{cor:qGEN to MAN}
\end{corollary}
\begin{proof} We need to show that the antisymmetric flow has a Hamiltonian structure. The proof becomes mostly straightforward after using Lemma~\ref{lem:pGEN to MAN} with $b(\rho):=\JJ(\rho)d\phi_\rho\tp d\E(\rho)$. From \eqref{eq:PsiMFT from PsiGEN}, since $\Psi_\GEN$ is a dissipation potential,
\begin{align*}
  d\phi_\rho d_\zeta\Psi^*_\MFT\big(\rho,F^\asym(\rho)\big)=d\phi_\rho d_\zeta\Psi^*_\GEN(\rho,0)+ d\phi_\rho \JJ(\rho)d\phi_\rho\tp d\E(\rho)=\hat\JJ(\rho)d\E(\rho).
\end{align*}
To show the second claim we calculate:
\begin{align*}
  \Psi^*_\MFT\big(\rho,\zeta+\theta d\phi_\rho\tp d\E(\rho)\big)
  &=\Psi_\GEN^*\big(\rho,\zeta-F^\asym(\rho)+\theta d\phi_\rho\tp d\E(\rho)\big)\\
  & -\Psi_\GEN^*\big(\rho,-F^\asym(\rho))\big)\\
  & +\big\langle \zeta+\theta d\phi_\rho\tp d\E(\rho),\JJ(\rho)d\phi_\rho\tp d\E(\rho)\big\rangle.
\end{align*} 
The first term on the right-hand side does not depend on $\theta$ by definition of quasi-GENERIC, the second term is constant in $\theta$, and the third term does not depend on $\theta$ because of the skewsymmetry of the Poisson structure $\JJ$.
\end{proof}

We stress that the previous corollary requires that $\L$ induces a GENERIC structure \emph{in flux space}, hence the drift $b(\rho)$ should be Hamiltonian in \emph{flux space}. This practically never occurs, see Remark~\ref{rem:cumflux inertia}. So although the result shows that flux-space GENERIC is stronger than MANERIC, the result in itself it rather useless. 

In the next result we slightly weaken the premise; we shall encounter both an example where this argument fails (Section~\ref{sec:Andersen}), and an example where it does apply (Section~\ref{sec:zero+det}).
\begin{theorem} If $\L$ induces a pre-GENERIC structure $(\Psi_\GEN,\V,b)$ with even $\Psi_\GEN,\Psi_\GEN^*$ and $\inf\V=0$, and $d\phi_\rho b(\rho)=\hat\JJ(\rho)d\E(\rho)$ for some Hamiltonian structure $(\hat\JJ,\E)$, then $\L$ induces the MANERIC structure $(\Psi_\MFT,\V,\hat\JJ,\E)$ at all points where $\Psi_\MFT,\Psi^*_\MFT$ given by \eqref{eq:PsiMFT from PsiGEN} are well-defined.
\label{th:pGEN to MAN}
\end{theorem}
\begin{proof} 
We only need to show that the antisymmetric flow is Hamiltonian. Using Lemma~\ref{lem:pGEN to MAN}:
\begin{align*}
	&d\phi_\rho d_\zeta\Psi^*_\MFT\big(\rho,F^\asym(\rho)\big)\\
	  &\,= d\phi_\rho d_\zeta \big\lbrack \Psi^*_\GEN(\rho,\zeta-F^\asym(\rho)) - \Psi^*_\GEN(\rho,-F^\asym(\rho)) + \langle \zeta,b(\rho)\rangle\big\rbrack_{\zeta=F^\asym(\rho)} \\
	  &\,= d\phi_\rho \big\lbrack d_\zeta \Psi^*_\GEN(\rho,0) + b(\rho)\big\rbrack = d\phi_\rho b(\rho)=\hat\JJ(\rho)d\E(\rho).
\end{align*}
\end{proof}

\subsection{Equation level: MANERIC to quasi-GENERIC}
\label{subsec:comparison eq level}

We now compare MANERIC to GENERIC on the \emph{equation level}, meaning that we focus on the dynamics encoded in $d\phi_\rho$ and $j^0(\rho)$ while disregarding any variational structure. Instrumental to the comparison will be a type of linearisation of the force-flux response relation $j^0(\rho)=d_\zeta\Psi^*(\rho,F(\rho))$. Although the construction is rather straightforward, we believe it to be interesting in its own right.
For linear evolution equations, a related idea can be found in \cite{Dietert2015,RengerSchindler2022}. The construction below is mostly taken from \cite{GaoLiu2022}, with some minor changes to make it more useful for the setting of this paper.

\begin{proposition} Let $\Psi_\MFT,\Psi_\MFT^*$ be convex dual dissipation potentials so that $\Psi^*_\MFT(\rho,\zeta)$ is strictly convex and twice (Gateaux) differentiable in $\zeta\in T_\rho^*\W$ for all $\rho\in\Z$. Futhermore, let $\Z\ni\rho\mapsto F^1(\rho),F^2(\rho)\in T_\rho^*\W$ be two force fields. Then the corresponding zero-cost flux can be rewritten as,
\begin{align}
  j^0(\rho)&:=d_\zeta\Psi^*_\MFT(\rho,F^1(\rho)+F^2(\rho))             \label{eq:before quadratisation}\\
    &= \KK(\rho)F^1(\rho) + d_\zeta\Psi^*_\MFT\big(\rho,F^2(\rho)\big) \label{eq:single quadratisation}\\
    &=\MM(\rho)F^1(\rho) + \MM(\rho)F^2(\rho),                         \label{eq:double quadratisation}
\end{align}
with the linear positive definite operators:
\begin{align}
  \KK(\rho)&:=\int_0^1\!d^2_\zeta\Psi^*_\MFT\big(\rho,\theta F^1(\rho)+F^2(\rho)\big)\,d\theta, \label{eq:quadratised operator}\\
  \MM(\rho)&:=\int_0^1\!d^2_\zeta\Psi^*_\MFT\big(\rho,\theta (F_1(\rho)+F_2(\rho)\big)\,d\theta \notag. 
\end{align}
\label{prop:linearised force structure}
\end{proposition}

\begin{proof}
The positive definiteness follows from the strict convexity of $\Psi^*_\MFT$. For the zero-cost flows,
\begin{align*}
  &d_\zeta\Psi^*_\MFT\big(\rho,F^1(\rho)+F^2(\rho)\big)-d_\zeta\Psi^*_\MFT\big(\rho,F^2(\rho)\big) \\
  &\qquad=\int_0^1\!\mfrac{d}{d\theta} d_\zeta\Psi^*_\MFT\big(\rho,\theta F^1(\rho)+F^2(\rho)\big)\,d\theta\\
  &\qquad=\KK(\rho) F^1(\rho),\\[-1em]
  &d_\zeta\Psi^*_\MFT\big(\rho,F^1(\rho)+F^2(\rho)\big)-\overbrace{d_\zeta\Psi^*_\MFT(\rho,0)}^{=0} \\
  &\qquad=\int_0^1\!\mfrac{d}{d\theta} d_\zeta\Psi^*_\MFT\big(\rho,\theta (F^1(\rho)+F^2(\rho))\big)\,d\theta\\
  &\qquad=\MM(\rho) (F^1(\rho)+F^2(\rho)).
\end{align*}
\end{proof}

For both constructions \eqref{eq:single quadratisation},\eqref{eq:double quadratisation} the force as well as the flux are decomposed into two components, even though for the original force-flux relation this was only true for quadratic dissipation potentials. The price to pay for this is that the natural variational structure is changed into a quadratic one. More particular, introducing the norms $\lVert j\rVert^2_{\KK(\rho)^{-1}}:=\sup_{\zeta\in T_\rho^*\W} 2\langle\zeta,j\rangle - \langle \zeta,\KK(\rho)\zeta\rangle$, and similar for $\lVert\zeta\rVert^2_{\MM(\rho)^{-1}}$, the cost functions that induce the force structures \eqref{eq:before quadratisation}, \eqref{eq:single quadratisation}, respectively \eqref{eq:double quadratisation} are uniquely given by:
\begin{align}
  \L(\rho,j)    &:=\Psi_\MFT(\rho,j)+\Psi^*_\MFT\big(\rho,F^1(\rho)+F^2(\rho)\big) - \langle F^1(\rho)+F^2(\rho),j\rangle,
  \label{eq:original L}\\
  \L_\KK(\rho,j)&:=\tfrac12 \big\lVert j - \KK(\rho)F^1(\rho) - d_\zeta\Psi^*_\MFT(\rho,F^2(\rho)) \big\rVert^2_{\KK(\rho)^{-1}},
  \label{eq:quadratised L K}\\
  \L_\MM(\rho,j)&:=\tfrac12 \big\lVert j-\MM(\rho)F^1(\rho) - \MM(\rho)F^2(\rho) \big\rVert^2_{\MM(\rho)^{-1}}.
  \label{eq:quadratised L M} 
\end{align}

In the context of the current paper we are mainly interested in taking $F^1=F^\sym$ and $F^2=F^\asym$ in construction~\eqref{eq:single quadratisation}. At first sight, the construction \emph{seems} to leave the decomposition of forces $F=F^\sym+F^\asym$ intact, and it is tempting to interpret $\KK(\rho)F^\sym(\rho), d_\zeta\Psi^*_\MFT(\rho,F^\asym(\rho))$ as the symmetric and antisymmetric parts of the \emph{flux}. However, since the variational structure is changed, the forces $F^\sym,F^\asym$ associated to the original cost $\L$ may no longer be the symmmetric and antisymmetric forces associated to the new cost function $\L_\KK$. 



The main theorem of this section states that MANERIC implies quasi-GENERIC, under a certain condition on the Hamiltonian structure $(\hat\JJ,\E)$. To understand this condition, let us assume that there is a differentiable quantity $\E^2$ such that $d\E^2(\rho)\in \Ker d\phi\tp_\rho$ for all $\rho\in\Z$. This means that $\E^2$ is conserved under any dynamics satisfying the continuity equation $\dot\rho(t)=d\phi_{\rho(t)} j(t)$, and in particular $\E^2$ is conserved under the (Hamiltonian) antisymmetric flow. In practice such conserved quantity can be the total mass, or in the context of chemical reactions, atom numbers~\cite{AndersonCraciunKurtz2010,MPPR2017}. For many Hamiltonian systems, there then exists an alternative Poisson structure such that $\hat\JJ(\rho)d\E(\rho)=\hat\JJ^2(\rho)d\E^2(\rho)$. We shall encounter two such Hamiltonian systems for the same equation in Subsection~\ref{eq:zero range Hamiltonian}. Under additional assumptions on the two Poisson structures $\hat\JJ$ and $\hat\JJ^2$, such systems are known as \emph{bi-Hamiltonian systems}
, see footnote~\ref{fn:bi-Hamiltonian}. We will not impose such additional compatibility assumption, but instead we simply assume that the Hamiltonian structure $(\hat\JJ,\E)$ is already in that rewritten form.

\begin{theorem} Let $\Psi_\MFT,\Psi^*_\MFT$ be convex dual dissipation potentials on $T\W$, $\JJ(\rho):T_\rho^*\Z\to T_\rho\Z$ be a Poisson structure, $\V,\E:\Z\to\RR$ be differentiable functionals with $\inf\V=0$. If the state-flux equations \eqref{eq:state-flux equations2} have the MANERIC structure $(\Psi_\MFT,F,\V,\hat\JJ,\E)$ and $d\E(\rho)\in\Ker(d\phi_\rho\tp)$ for all $\rho\in\Z$, then the state equation $\dot\rho(t)=d\phi_\rho j^0\big(\rho(t)\big)$ has the quasi-GENERIC structure~$(\hat\Psi_\GEN,\V,\hat\JJ,\E)$ 
\begin{align*}
  \hat\Psi_\GEN^*(\rho,\xi):=\mfrac12\big\langle d\phi_\rho\tp\xi,\KK(\rho)d\phi_\rho\tp\xi\big\rangle,
\end{align*}
with $\KK(\rho)$ given by \eqref{eq:quadratised operator}, $F^1=F^\sym, F^2=F^\asym$.
\label{th:MAN to qGEN}
\end{theorem}

\begin{proof} The proof is a direct application of the linearisation Proposition~\ref{prop:linearised force structure} above, the three-faces-of-the-second-law Corollary~\ref{cor:MFT 3 faces}, and the assumption $d\E(\rho)\in\Ker(d\phi_\rho\tp)$ implying:
\begin{align*}
  \hat\Psi_\GEN^*(\rho,\xi+\theta d\E(\rho)\big) \equiv \hat\Psi_\GEN^*(\rho,\xi\big).
\end{align*}
\end{proof}

In Subsection~\ref{subsec:zero-range qGEN} we shall use this argument to derive a quasi-GENERIC structure from a MANERIC structure.

\section{Example of LDP-induced MANERIC:\\ Zero-range process on a finite graph}
\label{sec:zero range}

The zero-range process~\cite[Sec.~2.3]{KipnisLandim1999} is often studied in connection with a hydrodynamic limit: on a lattice that is rescaled with $n$, yielding large deviations that induce quadratic dissipation, see~\cite{GessHeydecker2023TR}. By contrast, we study the zero-range process on a \emph{fixed finite} graph, because this scaling will yield non-quadratic dissipation potentials. 

The macroscopic fluctuation analysis of this model has already been carried out to a large extent in~\cite[Subsec.~5.1]{PattersonRengerSharma2024}. We continue the analysis here by showing that its large deviations induce a MANERIC structure. We then apply the results from Subsection~\ref{subsec:comparison eq level} to construct a quasi-GENERIC structure for the corresponding equation. As a by-product we obtain two altered cost functions, coming from the large deviations of two different microscopic systems.

\subsection{Microscopic model}
\label{subsec:zero range micro}

The zero-range process consists of $n\in\NN$ interacting particles $X_1(t),\hdots,X_n(t)$ that hop around on a finite graph $\X$. Zero range means that the rate with which a particle jumps from a site $x\in\X$ to another site depends on the number of particles at site $x$ only. The graph is equipped with an arbitrary ordering $<$, so that we can define the directed edges as $\X^2/2:=\{(x,y)\in\X\times\X:x<y\}$. We are interested in the empirical measure and cumulative empirical net flux, defined as, for $x\in\X$ respectively $(x,y)\in\X^2/2$,
\begin{align*}
  \rho\super{n}_x(t)&:=\frac1n\sum_{i=1}^n \mathds1_{X_i(t)}(x),\\
  W\super{n}_{xy}(t)&:=\frac1n\sum_{i=1}^n \sum_{\substack{s\in(0,T):\\ X_i(s^-)\neq X_i(s)}} \big( \mathds1_{(X_i(s^-),X_i(s))}(x,y)-\mathds1_{(X_i(s^-),X_i(s))}(y,x)\big).
\end{align*}
In other words, $\rho\super{n}_x(t)$ is the number of particles in state $x$ at time $t$ and $W\super{n}_{xy}(t)$ is the net number of particles that have jumped from $x$ to $y$ within the time span $(0,t)$.

To describe the precise dynamics, let $Q\in\RR^{\X\times\X}$ be a generator matrix of a Markov chain $X(t)$ on a finite graph $\X$. Assuming the chain is irreducible, it has a unique, coordinate-wise positive invariant measure $\pi\in\P(\X)$ such that $Q\tp\pi=0$ (after setting $Q_{xx}:=-\sum_{y\in\X} Q_{xy}$). In the case of independent copies, the rate for one particle to jump from $x$ to $y$ would be $n\rho_x\super{n} Q_{xy}$. For the zero-range process, we replace that rate by $\pi_x Q_{xy}\eta_x\big(\mfrac{\rho_x}{\pi_x}\big)$, introducing a family a functions $\eta_x:\lbrack0,\infty)\to\lbrack0,\infty), x\in\X$ that satisfy the following assumptions.
\begin{enumerate}[(i)]
\item $\eta_x$ is strictly increasing,
\item $\eta_x(0)=0$ and $\eta_x(1)=1$,
\item $\eta_x(z)=0$ is integrable near $a=0$.
\end{enumerate}
The condition $\eta_x(0)=0$ rules out negative concentrations. The condition $\eta(1)=1$ can be posed without loss of generality (see \cite[Subsec.~5.1]{PattersonRengerSharma2024}) and makes sure that $\pi$ is also the steady state for the limit equation~\eqref{eq:zero range limit eqns}. The strict monotonicity yields uniqueness of that steady state - among other facts. The integrability condition is a necessary and sufficient condition for the LDP to hold~\cite{AAPR2022}.

The pair $(\rho\super{n}(t),W\super{n}(t))$ is then a Markov process with generator:
\begin{align}
  &(\Q\super{n}f)(\rho,w)           \label{eq:zero range generator} \\
     &\qquad=n\sumxly \pi_x Q_{xy}\eta_x\big(\mfrac{\rho_x}{\pi_x}\big) \big\lbrack f(\rho-\tfrac1n\mathds1_x+\tfrac1n\mathds1_y,w+\tfrac1n\mathds1_{xy})-f(\rho,w)\big\rbrack \notag\\
     &\hspace{8em}+   \pi_yQ_{yx}\eta_y\big(\mfrac{\rho_y}{\pi_y}\big) \big\lbrack f(\rho-\tfrac1n\mathds1_y+\tfrac1n\mathds1_x,w-\tfrac1n\mathds1_{xy})-f(\rho,w)\big\rbrack,                   \notag
\end{align}
and the microscopic continuity equation $\rho\super{n}(t)=\rho^0-\Ddiv W\super{n}(t)$ holds almost surely by construction, where $\rho^0$ is the given initial condition.

\subsection{Macroscopic model and large deviations}
\label{subsec:zero range macro}

By the classic Kurtz convergence theorem~\cite{Kurtz1970a,EthierKurtz09}, the process $(\rho\super{n}(t),W\super{n}(t))$ converges (narrowly in Skorohod space) as $n\to\infty$ to the deterministic trajectory satisfying the limit equation:
\begin{align}
  \dot \rho_x(t)&=-\ddiv_x \dot w(t), &x\in\X,\notag\\ 
  \dot w_{xy}(t)&=j^0_{xy}\big(\rho(t)\big):=\pi_x Q_{xy} \eta_x\big(\mfrac{\rho_x(t)}{\pi_x}\big) - \pi_y Q_{yx} \eta_x\big(\mfrac{\rho_y(t)}{\pi_y}\big),
  &(x,y)\in\X^2/2,
\label{eq:zero range limit eqns}
\end{align}
where the discrete divergence and its adjoint (the discrete gradient) are
\begin{align}
  &\ddiv_x j:=\sum_{y\in\X:y> x}j_{xy} -\sum_{y\in\X:y<x} j_{yx},  &x\in\X,
  \label{eq:discrete divergence}\\
  &\dgrad_{xy} \xi=\xi_y - \xi_x, &(x,y)\in\X^2/2.\notag
\end{align}
The equations \eqref{eq:zero range limit eqns} are indeed of the form~\eqref{eq:limit evolution} if we set the net flux to be the derivative of the cumulative net flux $j(t):=\dot w$, and the continuity operator to $d\phi_\rho:=-\ddiv$, see also \eqref{eq:zero range cont op} below.

Since we assumed the integrability condition on each of the $\eta_x$-functions, the pathwise LDP~\eqref{eq:framework LDP} corresponding to $n\to\infty$ Kurtz limit holds~\cite{AAPR2022}, with \emph{non-quadratic} cost function:
\begin{align}
  \L_s(\rho,j) &:=
    \sumxly \inf_{j_{xy}^+\geq0}
    \bigl[   s\big(j^+_{xy} \mid \pi_xQ_{xy} \eta_x(\tfrac{\rho_x}{\pi_x})\big) + s\big(j^+_{xy}-j_{xy} \mid \pi_y Q_{yx}\eta_y(\tfrac{\rho_y}{\pi_y} \big)\bigr],
\label{eq:zero range L}\\
  s(a\mid b) &:=
    \begin{cases}
      a\log\frac{a}{b}-a+b, &a,b>0,\\
      b,                    &a=0,\\
      \infty,               &b\leq0,a>0 \text{ or  } a<0, 
    \end{cases}\label{def:entropy-s}
\end{align}
Note that indeed $\L_s(\rho,j)$ is non-negative and minimised by \eqref{eq:zero range limit eqns}. The infimum in \eqref{eq:zero range L} appears from a contraction principle~\cite[Thm.~4.2.1]{DemboZeitouni09} applied to the large-deviation rate functional for one-way rather then net fluxes. For future reference we also introduce the convex dual \eqref{eq:H dual L} of $\L$:
\begin{equation}
  \H_s(\rho,\zeta):=\sumxly \big[ \pi_x Q_{xy}\eta_x(\tfrac{\rho_x}{\pi_x})\big(e^{\zeta_{xy}}-1\big) + \pi_y Q_{yx}\eta_y(\tfrac{\rho_y}{\pi_y})\big(e^{-\zeta_{xy}}-1\big)\big].
\label{eq:zero range H}
\end{equation}

\subsection{Geometry}
\label{subsec:zero range geometry}

We first describe how the model fits in the geometric framework of Subsection~\ref{subsec:framework}, using the notions taken from~\cite[Example~2.2]{PattersonRengerSharma2024}. The state space $\Z$ is the vector space of densities on $\X$ with total mass one, with corresponding (co)tangent space and Euclidean pairing between them:
\begin{align*}
  \Z        &=\{ {\textstyle \rho\in\RR^\X:\sum_{x\in\X} \rho_x=1 } \},
  &&_{T_\rho^*\Z}\langle \xi,u\rangle_{T_\rho\Z}=\xi\cdot u,\\
  T_\rho\Z  &=\{ {\textstyle   u \in\RR^\X:\sum_{x\in\X} u_x=0 } \},
  &&
  T^*_\rho\Z=\RR^\X\,\mathrm{mod}\,(1,\hdots,1).
\end{align*}
Observe that the pairing is indeed independent of the representative of the equivalence class $\xi\in T_\rho^*\Z$. For the fluxes we simply take the flat space:
\begin{align*}
  \W=T_\rho\W=T_\rho^*\W =\RR^{\X^2/2},
  &&_{T_\rho^*\W}\langle \zeta,j\rangle_{T_\rho\W}=\zeta\cdot j.
\end{align*}
Finally, the operator $\phi:\W\to\Z$ encodes the time-integrated continuity equation, for some fixed reference state or initial condition $\rho(0)$,
\begin{align}
  \phi[w]:=\rho(0)-\ddiv w, && d\phi_\rho=-\ddiv, &&  d\phi_\rho\tp=\dgrad.
\label{eq:zero range cont op}
\end{align}
Now that the spaces $\Z,\W$ and continuity operator $\phi$ are defined, we are ready to analyse the large-deviation cost \eqref{eq:zero range L} and corresponding zero-cost flux~\eqref{eq:zero range limit eqns}.

\subsection{Macroscopic Fluctuation Theory}

The force structure $(\Psi_\MFT,F)$ induced by $\L_s$ is easily found using~\eqref{eq:zero range H} together with the explicit formulas in Proposition~\ref{prop:force structure}, see also \cite[Subsec.~5.1]{PattersonRengerSharma2024}:
\begin{align}
  \Psi^*_\MFT(\rho,\zeta)&=2\sumxly \sqrt{\pi_x Q_{xy}\eta_x(\tfrac{\rho_x}{\pi_x}) \pi_y Q_{yx}\eta_y(\tfrac{\rho_y}{\pi_y}) } \big(\cosh(\zeta_{xy})-1\big),
  \label{eq:zero-range Psi*}\\
  \Psi_\MFT(\rho,j)&=2\sumxly \sqrt{\rho_x Q_{xy}\rho_y Q_{yx}}\big(\cosh^*(\mfrac{j_{xy}}{2\sqrt{\rho_x Q_{xy}\rho_y Q_{yx}}})+1\big),\notag\\
  F_{xy}(\rho)&=\frac12\log\frac{\pi_x Q_{xy}\eta_x(\rho_x/\pi_x)}{\pi_y Q_{yx}\eta_y(\rho_y/\pi_y)}.
  \label{eq:zero-range force}
\end{align}

Next we identify the quasipotential $\V$ that solves the Hamilton-Jacobi equation~\eqref{eq:HJE}. This has been shown for the zero-range process in~\cite[Prop. 5.1]{PattersonRengerSharma2024} with:
\begin{align}
  \V(\rho):=
  \begin{cases}
    \sum_{x\in\X} \int_0^{\rho_x}\!\log \eta_x\big(\mfrac{a}{\pi_x}\big)\,da,  &\rho\in\P(\X),\\
    \infty,                                                                    &\text{otherwise}.
  \end{cases}
\label{eq:zero-range QP}
\end{align}
As noted in the same paper, this quasipotential can also be found using the explicitly known invariant measure $\Pi\super{n}$ of the zero-range process~\cite[Prop.~3.2]{KipnisLandim1999}, calculate the corresponding large deviations, and then use our Proposition~\ref{prop:MFT micro macro}\emph{(i)} to find the same quasipotential $\V$, see also \cite[Sec.~4.1]{GabrielliRenger2020}. Observe that the integrals are well-defined due to the assumptions on $\eta_x$.
With this quasipotential $\V$, the symmetric and antisymmetric components~\eqref{eq:Fsym Fasym} of the force are then:
\begin{align}
  F^\sym_{xy}(\rho)=\frac12\log\frac{\eta_x(\rho_x/\pi_x)}{\eta_y(\rho_y/\pi_y)}, \hspace{1.5em}
  F^\asym_{xy}(\rho)=\frac12\log\frac{\pi_x Q_{xy}}{\pi_y Q_{yx}}.
\label{eq:zero-range forces}
\end{align}
And hence the full dynamics~\eqref{eq:MFT general dynamics} is driven by a combination of both forces.

The symmetric flow~\eqref{eq:MFT symmetric flow} is obtained when $F^\asym(\rho)=0$. This means that the process $(\rho\super{n}(t),W\super{n}(t))$ is in detailed balance, which is equivalent to
\begin{align*}
    \pi_xQ_{xy}=\pi_y Q_{yx} && \text{for all } x,y \in \X.
\end{align*}
In this case $\L_s$ induces the gradient flow of the quasipotential $\V$ as discussed in Subsection~\ref{subsec:MFT symmetric flow}. 

In the next section we focus on the antisymmetric flow.

\subsection{Antisymmetric flow and MANERIC}
\label{eq:zero range Hamiltonian}

According to Definition~\ref{def:MANERIC}, $\L_s$ induces a MANERIC structure if the antisymmetric flow~\eqref{eq:MFT antisymm flow} has a Hamiltonian structure. After filling in $\Psi^*_\MFT$ and $F^\asym$, the antisymmetric flow of the zero-range process becomes
\begin{equation}
  \dot\rho_x(t) = \sum_{\substack{y\in\X\\y\neq x}} A_{xy} \sqrt{\pi_x\eta_x\big(\tfrac{\rho_x(t)}{\pi_x}\big)\pi_y\eta_y\big(\tfrac{\rho_y(t)}{\pi_y}\big)}, \quad \text{with} \quad A_{xy}:= Q_{yx} \sqrt{\mfrac{\pi_y}{\pi_x}} - Q_{xy}\sqrt{\mfrac{\pi_x}{\pi_y}}. 
\label{eq:zero-range antisym ODE}
\end{equation}
It was shown in ~\cite[Subsec.~4]{PattersonRengerSharma2024} that in the special $\eta_x\equiv\mathrm{id}$ -- corresponding to independent particles -- this equation has a Hamiltonian structure, and in the more general case~\eqref{eq:zero-range antisym ODE}, $\dot\rho(t)=\hat\JJ(\rho(t))d\E(\rho(t))$ for some energy $\E$ and skewsymmetric operator $\hat\JJ(\rho)$, but the Jacobi identity may not hold anymore. We continue that analysis here to derive a different structure $(\hat\JJ,\E)$ that does satisfy the Jacobi identity.

\paragraph{Step 1.} Inspired by the geometric argument for the independent particles case~\cite[Subsec.~4]{PattersonRengerSharma2024}, the analysis becomes easier after transforming to the new variable $\omega_x(t):=g_x(\rho_x(t)):=\int_0^{\rho_x(t)}\!\big(\pi_x\eta_x(a/\pi_x)\big)^{-1/2}\,da$, so that \eqref{eq:zero-range antisym ODE} becomes
\begin{align}
  \dot\omega_x(t)&=\sum_{\substack{y\in\X:\\y\neq x}} A_{xy}\, (g_y^{-1})'(\omega_y(t)).
\label{eq:zero-range antisym omega}
\end{align}

\paragraph{Step 2.} Restricting to the case $\lvert\X\rvert=3$, we claim that equation~\eqref{eq:zero-range antisym omega} has \emph{two} distinct conserved quantities, namely
\begin{align*}
  \tilde\E^1(\omega):=-A_{23}\omega_1 - A_{31}\omega_2 - A_{12}\omega_3, &&
  \tilde\E^2(\omega):=\sum_{x=1,2,3} g_x^{-1}(\omega_x),
\end{align*}
using tildes to signify quantities defined on $\omega$.
The first function $\tilde\E^2(\omega)=\sum_{x=1,2,3} \rho_x$ is simply the total mass, by construction conserved under the antisymmetric flow $\dot\rho(t)=-\ddiv d_\zeta\Psi^*_\MFT(\rho(t),F^\asym(\rho(t)))$. To see that $\E^1$ is conserved, $\grad\tilde\E^1(\omega)$ must be orthogonal to the right-hand side of the ODE~\eqref{eq:zero-range antisym omega} (for brevity writing $\X=\ZZ\mod 3$):
\begin{align*}
  &\sum_{x=1,2,3} \sum_{\substack{y=1,2,3,\\ y\neq x}} -A_{x+1,x+2} A_{xy}\, (g_y^{-1})'(\omega_y)\\
  &\quad \sum_{y=1,2,3} (g_y^{-1})'(\omega_y) \underbrace{(A_{y+2,y}A_{y+1,y}+A_{y,y+1}A_{y+2,y})}_{=0 \text{ (skew symmetry of $A$)}} =0.
\end{align*}
Note that at this stage $\tilde\E^1,\tilde\E^2$ may be multiplied by arbitrary constants, but we have already chosen those constants so that the next step works out well.

\paragraph{Step 3.}  The reason to restrict to 3 dimensions is that the Jacobi identity corresponding to a Hamiltonian structure ($\tilde\JJ,\tilde\E)$ implies that $\tilde\JJ(\omega)\grad\tilde\E(\omega)=\grad\tilde\E(\omega)\times\grad\tilde{\mathcal{C}}(\omega)$ for some conserved ``Casimir'' function $\tilde{\mathcal{C}}$ \cite{EsenGhoseGuha2016}. Hence we can use the two conserved quantities $\tilde\E^1,\tilde\E^2$ to construct:
\begin{align*}
  \tilde\JJ^1(\omega)&:=\cdot \times \grad\tilde\E^2(\omega) =
    \begin{bmatrix} 0                      & (g_3^{-1})'(\omega_3)  & -(g_2^{-1})'(\omega_2) \\
                    -(g_3^{-1})'(\omega_3) &    0                   &  (g_1^{-1})'(\omega_1) \\
                    (g_2^{-1})'(\omega_2)  & -(g_1^{-1})'(\omega_1) &  0                     \\
    \end{bmatrix},\\
  \tilde\JJ^2(\omega)&:= \grad\tilde\E^1(\omega)\times\cdot =
    \begin{bmatrix} 0       & A_{12}  & -A_{31}  \\
                    -A_{12} & 0       & A_{23} \\
                     A_{31} & -A_{23} & 0       \\
    \end{bmatrix}
    =A.
\end{align*}
One then sees that
\begin{align*}
  \tilde\JJ^1(\omega)\grad\tilde\E^1(\omega)&=\tilde\JJ^2(\omega)\grad\tilde\E^2(\omega)
  =\grad\tilde\E^1(\omega)\times\grad\tilde\E^2(\omega)\\
  &=\begin{bmatrix} -A_{31} (g_3^{-1})'(\omega_3) + A_{12}(g_2^{-1})'(\omega_2)\\
                   -A_{12} (g_1^{-1})'(\omega_1) + A_{23}(g_3^{-1})'(\omega_3)\\
                   -A_{23} (g_2^{-1})'(\omega_2) + A_{31}(g_1^{-1})'(\omega_1)
   \end{bmatrix},
\end{align*}
which is precisely the right-hand side of \eqref{eq:zero-range antisym omega}.  Since the Jacobi identity for $\tilde\JJ^1,\tilde\JJ^2$ hold by construction, we have thus found that in three dimensions, the equation~\eqref{eq:zero-range antisym omega} has two distinct Hamiltonian structures\footnote{This would be called a ``bi-Hamiltonian system'' whenever linear combinations of the two Poisson structures are again a Poisson structure. We do not pursue this further since we do not need it. \label{fn:bi-Hamiltonian}} $(\tilde\JJ^1,\tilde\E^1)$ and $ (\tilde\JJ^2,\tilde\E^2)$.

\paragraph{Step 4.} We now transform the structures back to the state space variables $\rho$. If allowing the slight abuse of notation $g:\RR^3\to\RR^3, g_x(\rho)=g_x(\rho_x)$ for the transformation from step 1, then the conserved quantities and Poisson structures are transformed back through $\E^{1,2}(\rho):=(\tilde\E^{1,2}\circ g)(\rho)$ and $(\hat\JJ^{1,2}\circ g)(\omega):=(g^{-1})'(\omega)\tilde\JJ^{1,2}(\omega)(g^{-1})'(\omega)$, yielding
\begin{align*}
  \E^1(\rho)=-A_{23}g_1(\rho_1) - A_{31}g_2(\rho_2) - A_{12}g_3(\rho_3), &&
  \E^2(\rho)=\sum_{x=1,2,3} \rho_x,
\end{align*}
and
\begin{align*}
  &\hat\JJ^1(\rho)=\sqrt{\pi_1\eta_1(\tfrac{\rho_1}{\pi_1}) \pi_2\eta_2(\tfrac{\rho_2}{\pi_2}) \pi_3\eta_3(\tfrac{\rho_3}{\pi_3})} \begin{bmatrix} 0 & 1 & -1\\ -1 & 0 & 1 \\ 1 & -1 & 0\end{bmatrix},\\
  &\hat\JJ^2(\rho)=\\
  &\begin{bmatrix}
    0 & \sqrt{\pi_1\eta_1(\tfrac{\rho_1}{\pi_1}) \pi_2\eta_2(\tfrac{\rho_2}{\pi_2})} A_{12} & \sqrt{\pi_1\eta_1(\tfrac{\rho_1}{\pi_1}) \pi_3\eta_3(\tfrac{\rho_3}{\pi_3})} A_{13}\\
    -\sqrt{\pi_1\eta_1(\tfrac{\rho_1}{\pi_1}) \pi_2\eta_2(\tfrac{\rho_2}{\pi_2})} A_{12} & 0 & \sqrt{\pi_2\eta_2(\tfrac{\rho_2}{\pi_2}) \pi_3\eta_3(\tfrac{\rho_3}{\pi_3})} A_{23} \\
    -\sqrt{\pi_1\eta_1(\tfrac{\rho_1}{\pi_1}) \pi_3\eta_3(\tfrac{\rho_3}{\pi_3})} A_{13} & -\sqrt{\pi_2\eta_2(\tfrac{\rho_2}{\pi_2}) \pi_3\eta_3(\tfrac{\rho_3}{\pi_3})} A_{23} & 0
  \end{bmatrix}.
\end{align*}
Then the untransformed antisymmetric flow~\eqref{eq:zero-range antisym ODE} has the two Hamiltonian structures $(\JJ^1,\E^1), (\JJ^2,\E^2)$, where the Jacobi identities for $\JJ^1,\JJ^2$ are inherited from the Jacobi identities for $\tilde\JJ^1,\tilde\JJ^2$.

\paragraph{Step 5.} We now generalise to arbitrary dimensions. This is difficult for the first Hamiltonian structure $(\hat\JJ^1,\E^1)$, but easy for the second structure $(\hat\JJ^2,\E^2)$:
\begin{align}
  \E(\rho):=\sum_{x\in\X} \rho_x,&&
  \hat\JJ_{xy}(\rho):= \sqrt{\pi_x\eta_x(\tfrac{\rho_x}{\pi_x}) \pi_y\eta_y(\tfrac{\rho_y}{\pi_y})} A_{xy},
  &&x,y\in\X.
\label{eq:zero-range HS}
\end{align}
We could also have first generalised the transformed structure $\tilde\JJ(\omega)=A$ to arbitrary dimensions. Since $A$ is skew symmetric and does not depend on $\rho$, $\tilde\JJ$ automatically satisfies the Jacobi identity, and so after transforming back, $\hat\JJ$ also satisfies the Jacobi identity.

We conclude that in arbitrary dimensions, the antisymmetric flow~\eqref{eq:zero-range antisym ODE} has the Hamiltonian structure $(\hat\JJ,\E)$, and so
\begin{quote}
  The cost $\L_s$ defined in~\eqref{eq:zero range L} \textbf{induces the MANERIC structure} $(\Psi_\MFT,F,\V,\hat\JJ,\E)$,
with dissipation potential $\Psi_\MFT$, force $F$ and Hamiltonian structure $(\hat\JJ,\E)$ given by \eqref{eq:zero-range force} and \eqref{eq:zero-range HS}. 
\end{quote}
Moreover, the MANERIC structure has the additional property that the Hamiltonian system is driven by the total mass, which is conserved by $\phi$. This is essential to transform to quasi-GENERIC in the next subsection.

\subsection{MANERIC to quasi-GENERIC}
\label{subsec:zero-range qGEN}

By Proposition~\ref{prop:preGEN} the cost function $\L_s$ from~\eqref{eq:zero range L} for the zero-range process does not induce a (pre/quasi-)GENERIC structure. This becomes a different story if we ignore the variational structure that comes with $\L_s$ and focus on the full dynamics~\eqref{eq:zero range limit eqns} contracted to the state space:
\begin{equation}
  \dot\rho_x(t)=-\ddiv_x j^0(\rho(t))=\sum_{y\neq x}\Big\lbrack \pi_y Q_{yx}\eta_y\big(\tfrac{\rho_y}{\pi_y}\big) - \pi_x Q_{xy}\eta_x\big(\tfrac{\rho_x}{\pi_x}\big) \Big\rbrack.
\label{eq:zero-range contracted equation}
\end{equation}

For the MANERIC structure derived in the previous subsection with Hamiltonian structure~\eqref{eq:zero-range HS} we check that indeed $\grad\E(\rho)\in\Ker(\dgrad)$, that is, $\dgrad(1,\hdots,1)=0$. Hence we can apply Theorem~\ref{th:MAN to qGEN} to obtain a quasi-GENERIC structure for equation~\eqref{eq:zero-range contracted equation}.

The only thing that needs to be done is to calculate the linear, positive definite operator $\KK(\rho):T_\rho^*\W\to T_\rho\W$ from \eqref{eq:quadratised operator}, $F^1=F^\sym, F^2=F^\asym$. Abbreviating $\eta_x=\eta_x\big(\tfrac{\rho_x}{\pi_x}\big)$, this yields,
\begin{align}
  \KK_{xy,x'y'}(\rho)=\frac{\pi_x Q_{xy}\big(\eta_x-\sqrt{\eta_x\eta_y}\big) - \pi_y Q_{yx} \big( \eta_y-\sqrt{\eta_x\eta_y}\big)}{1/2 \log\big(\eta_x/\eta_y\big)}\mathds1_{xy=x'y'}.
\label{eq:zero-range K}
\end{align}
As shown in Proposition~\ref{prop:linearised force structure}, the positive definiteness of $\KK$ follows from the convexity of $\Psi_\MFT^*$. Contracted to the state space, this operator $\hat\KK(\rho):T_\rho^*\Z\to T_\rho\Z$ and the corresponding quadratic dual dissipation potential $\hat\Psi_\GEN:T\Z\to\lbrack0,\infty\rbrack$ are (similar to Remark~\ref{rem:gradient flow flux form} and \eqref{eq:contracted JJ}):
\begin{align}
  \hat\KK_{x,y}(\rho)&:= (\ddiv\KK(\rho)\dgrad)_{xy}=
  \begin{cases}
    2 \sum_{z\neq x} \KK_{xz,xz}(\rho), &x=y,\\
   -2 \KK_{xy,xy}(\rho),                &x\neq y,
  \end{cases}\notag\\
  \hat\Psi^*_\GEN(\rho,\xi)&:=\tfrac12\langle \dgrad\xi,\KK(\rho)\dgrad\xi\rangle=\tfrac12\langle \xi,\hat\KK(\rho)\xi\rangle.
\label{eq:zero-rangle qua Psis}
\end{align}
Thus by Theorem~\ref{th:MAN to qGEN} and Definition~\ref{def:rho GEN},
\begin{quote}
  The evolution equation~\eqref{eq:zero-range contracted equation} \textbf{has the quasi-GENERIC structure}~$(\hat\Psi_\GEN,\V,\hat\JJ,\E)$,
  with contracted dissipation potential $\hat\Psi_\GEN$, free energy $\V$ and Hamiltonian structure $(\hat\JJ,\E)$ given by ~\eqref{eq:zero-rangle qua Psis}, \eqref{eq:zero-range QP} and \eqref{eq:zero-range HS}.
\end{quote}
Recall that this statement means that 
\begin{enumerate}[(i)]
  \item $u^0(\rho):=-\ddiv j^0(\rho)= -\tfrac12\KK(\rho)\grad\V(\rho)\big) + \hat\JJ(\rho)\grad\E(\rho)$,
  \item $\big\langle d\V(\rho),u^0(\rho)\big\rangle\leq0$,
  \item $\hat\Psi^*_\GEN(\rho,\xi+\theta \grad\E(\rho))\equiv\hat\Psi^*(\rho,\xi)$.
\end{enumerate}
The argument of Theorem~\ref{th:MAN to qGEN} is that the first statement holds by construction, the second statement by the second law Corollary~\ref{cor:MFT 3 faces} applied to $v_0$, and the last statement since $\dgrad\grad\E(\rho)=0$.

\subsection{Two alternative microscopic models}

\paragraph{$\KK$-structure.} In order to pass from MANERIC to quasi-GENERIC in the previous subsection, we essentially replaced the original cost \eqref{eq:original L} by the quadratised version \eqref{eq:quadratised L K}. In fact, this cost function $\L_\KK$ corresponds to a different microscopic model, driven by white noise (similar to the reverse-engineering argument of \cite{ReinaZimmer2015}):
\begin{align*}
  d\rho\super{n}(t)=-\ddiv dW\super{n}(t), &&
  dW\super{n}(t)=j^0(\rho\super{n}(t))\,dt + \frac{1}{\sqrt{n}}\KK(\rho) dB(t),
\end{align*}
where $j^0$ is the zero-cost flux~\eqref{eq:zero range limit eqns} and $(B_{xy}(t))_{(x,y)\in\X^2/2}$ are independent Brownian motions on the positively directed edges of the graph. The model is slightly reminiscent of the classic Ginzburg-Landau model~\cite{Spohn1985}, which is also driven by Brownian motions on its edges. Indeed, by Freidlin-Wentzell techniques~\cite{Freidlin2012} one obtains that that the above model $(\rho\super{n}(t),W\super{n}(t))$ satisfies the LDP with quadratic cost function $\L_\KK$, and applying the arguments from Section~\ref{sec:GENERIC} to $\L_\KK$ yields the quasi-GENERIC structure~$(\hat\Psi_\GEN,\V,\hat\JJ,\E)$ from Subsection~\ref{subsec:zero-range qGEN}.

One might argue that the Onsager structure $\KK$ from \eqref{eq:zero-range K} that we use here is not a very natural one. Indeed, the denumerator $1/2 \log(\eta_x/\eta_y)$ appears precisely so that it cancels out when applied to the symmetric force~\eqref{eq:zero-range forces}.

\paragraph{$\MM$-structure.} Alternatively, we can replace the nonquadratic cost~\eqref{eq:original L} by the other quadratised version \eqref{eq:quadratised L M}, which is the large-deviation cost of the following microscopic model:
\begin{align*}
  d\rho\super{n}(t)=-\ddiv dW\super{n}(t), &&
  dW\super{n}(t)=j^0(\rho\super{n}(t))\,dt + \frac{1}{\sqrt{n}}\MM(\rho) dB(t),
\end{align*}
with positive definite operator
\begin{align}
  \MM_{xy,x'y'}(\rho)&=2\sumxly \sqrt{\pi_x Q_{xy}\eta_x(\tfrac{\rho_x}{\pi_x}) \pi_y Q_{yx}\eta_y(\tfrac{\rho_y}{\pi_y}) } \frac{\sinh(F_{xy}(\rho))}{F_{xy}(\rho)}\,\mathds1_{xy=x'y'}\notag\\
  &=2\Lambda\Big( \pi_x Q_{xy}\eta_x\big(\tfrac{\rho_x}{\pi_x}\big),\pi_y Q_{yx}\eta_y\big(\tfrac{\rho_y}{\pi_y}\big) \Big) \,\mathds1_{xy=x'y'},
\label{eq:zero-range MM}
\end{align}
and \emph{logarithmic mean} $\min\{u,v\}\leq\Lambda(u,v)\leq \max\{u,v\}$ given by
\begin{equation*}
    \Lambda(u,v):=\begin{cases} \frac{u-v}{\log u - \log v}, &u,v>0, u\neq v,\\
                                u,                           &u,v>0, u=v,\\
                                0,                           &u=0 \text{ or } v=0.
                  \end{cases}
\end{equation*}
Although the new cost function $\L_\MM$ is no longer compatible with the quasi-GENERIC structure~$(\hat\Psi_\GEN,\V,\hat\JJ,\E)$, it is interesting for other reasons. Apart from being a seemingly more natural structure, it coincides with the gradient structure introduced in~\cite{Mielke2011GF} and \cite{Maas2011}. The variational structure $\L_\MM$ is a thorough generalisation of those works, in that we allow for zero-range interaction through the functions $\eta_x$, as well as non-dissipative effects. 

It would thus be interesting to apply the MFT analysis to this new cost $\L_\MM$. However, due to the non-equilibrium nature, it turns out to be rather challenging to find the quasipotential. We show in Appendix~\ref{sec:M-structure} that in the detailed balance setting, the quasipotential coincides with \eqref{eq:zero-range QP}, and that away from detailed balance it can \emph{not} be the same function. The explicit solution to the Hamilton-Jacobi equation remains an open question.

\section{Example of LDP-induced pre-GENERIC:\\
Anderson Thermostat}
\label{sec:Andersen}


The previous Section~\ref{sec:zero range} provided an example for the theory of Subsection~\ref{subsec:comparison eq level}, that is: from an LDP-induced MANERIC structure to a quasi-GENERIC structure for the evolution equation. The current and next section aim to provide examples for the theory of Subsection~\ref{subsec:comparison LDP level}, that is: from an LDP-induced quasi/pre-GENERIC structure to an LDP-induced MANERIC structure, while keeping the variational structure intact.

As argued throughout this paper, a (quasi)-GENERIC structure in flux space usually does not exist, which is also true for the Andersen thermostat. Nevertheless, this model is often used as a prototypical example of large-deviation-based GENERIC~\cite{KLMP2018phys,KLMP2020math,DuongOttobre2023}, albeit it not in flux space. We shall see that the theory of Subsection~\ref{subsec:comparison LDP level} does not apply, meaning that the large deviations do not induce a MANERIC structure. We include this model anyhow, to illustrate why the MANERIC derivation breaks down.

\subsection{Microscopic model}

The Andersen Thermostat can be seen as a kinetic Monte Carlo method with resampling in order to make the dynamics ergodic. For simplicity we restrict to non-interacting particles on the $d$-dimensional torus $\TT^d$. See \cite{KLMP2020math} for the model with pair interaction.

More precisely, the model consists of particles with positions $X_i(t)\in\TT^d$ and velocities $V_i(t)\in\RR^d$. Each particle undergoes a deterministic motion and a random resampling of the velocities. It is therefore natural to distinguish the deterministic flux from the resampling flux. The empirical measure will then be almost surely coupled to the two fluxes (introduced below) via the continuity equation:
\begin{align*}
  \frac1n\sum_{i=1}^n \delta_{(X_i(t),V_i(t))}=:\rho\super{n}(t) =\rho\super{n}(0)-\div W\super{n}_\Det(t) -\ddiv W\super{n}_\Res(t),
\end{align*}
where the continuous divergence $\div$ acts on both coordinates $(x,v)$ and the discrete divergence is similar to~\eqref{eq:discrete divergence}:
\begin{equation}
  (\ddiv j_\Res)(dx\,dv):=\int_{v'\in\RR^d:v'> v} j_\Res(dx\,dv\,dv') -\int_{v'\in\X:v'<v} j_\Res(dx\,dv'\,dv).
\label{eq:Andersen ddiv}
\end{equation}

The resampling means that each particle $(X_i(t),V_i(t))$ independently changes its velocity according a Poisson point process with fixed intensity measure $\mu \in\M(\RR^d)$ (to be specified later). To define the corresponding (cumulative) resampling flux, we mimick the structure for the zero-range model from Section~\ref{sec:zero range} by equipping the velocity space $\RR^d$ with an (arbitrary) ordering $<$, and again write by a slight abuse of notation $\RR^{2d}/2:=\{(v,v')\in \RR^{2d}:v<v'\}$. The resampling flux is then defined as a measure on the set $\TT^d\times\RR^{2d}/2$:
\begin{multline*}
  W\super{n}_\Res(t,dx\,dv\,dv'):=\frac1n\sum_{i=1}^n \sum_{\substack{s\in(0,T):\\ V_i(s^-)\neq V_i(s)}} \Big\lbrack \delta_{(X_i(s^-),V_i(s^-),V_i(s))}(dx\,dv\,dv') \\
   - \delta_{(X_i(s^-),V_i(s),V_i(s^-))}(dx\,dv\,dv') \Big\rbrack,
\end{multline*}
The (cumulative) deterministic flux is driven by the velocity and a given potential $U\in C^1(\TT^d)$, yielding as a measure on the set $\{(x,v):x\in\TT^d,v\in \RR^d\}$,
\begin{align*}
  W\super{n}_\Det(t,dx\,dv) &= \int_0^t\!\rho\super{n}(s,dx\,dv)\begin{bmatrix} v\\-\grad_x U(x)\end{bmatrix} \,ds.
\end{align*}

\subsection{Macroscopic model and large deviations}

The random trajectories $(\rho\super{n}(t),W\super{n}_\Det(t),W\super{n}_\Res(t))$ converge as $n\to\infty$ to the coupled equations:
\begin{align*}
  &\dot\rho(t)=-\div \dot w_\Det(t) -\ddiv \dot w_\Res(t), \\
  &\dot w_\Det(t)=j^0_\Det(\rho(t)), \qquad \dot w_\Res(t)=j^0_\Res(\rho(t)), 
\end{align*}
with zero-cost flows:
\begin{align*}  
  j_\Det^0(\rho)(x,v)&:=\rho(x,v)\begin{bmatrix} v\\-\grad_x U(x)\end{bmatrix},\\
  j_\Res^0(\rho)(dx\,dv\,dv')&:= \underbrace{\mu(dv')\rho(dx\,dv)}_{=:(\rho\otimes\mu)(dx\,dv\,dv')} - \underbrace{\mu(dv)\rho(dx\,dv')}_{=:(\mu\otimes\rho)(dx\,dv\,dv')}.
\end{align*}

The LDP corresponding to this limit is described by the cost function:
\begin{align}
  \L(\rho,j) &:= \L_\Res(\rho,j_\Res) + \L_\Det(\rho,j_\Det), \label{eq:AT cost} \\
  \L_\Det(\rho,j_\Det) &:=\begin{cases} 0, &j_\Det=j^0_\Det(\rho),\\
                                    \infty, &\text{otherwise},
                          \end{cases} \label{eq:Andersen L}\\
  \L_\Res(\rho,j_\Res) &:= \inf_{\substack{j^+,j^-\in \M_{\geq0}(\TT^d\times\RR^{2d})\\ j_\Res(x,v,v') = j^+(x,v,v')-j^-(x,v',v)}} S(j^+\mid \rho\otimes\mu) + S(j^-\mid  \mu\otimes \rho),\notag\\
  S(j\mid \nu) &:= 
    \begin{cases}
      \iiint_{\TT^d\times\RR^{2d}/2}\,s\big(\frac{dj}{d\nu}\mid 1) \,dj,  &j\ll \nu, \qquad\text{ (recall $s$ from \eqref{def:entropy-s}) }\\
      \infty,                                            &\text{otherwise}.
    \end{cases}\notag
\end{align}
In addition, the continuity equation $\dot\rho(t)=-\div j_\Det(t)-\ddiv j_\Res(t)$ holds, otherwise the large-deviation cost is set to $\infty$.

\subsection{Geometry}

For the state space we use the infinite-dimensional version of the structure for the zero-range process. Let $\M(\RR^n)$ denote the space of signed measures on $\RR^n$ and $\M_\theta(\RR^n)$ signed measures with total mass $\int\!d\rho=\theta$, and set:
\begin{align*}
  \Z        &=\M_1(\TT^d\times\RR^d),
  &&_{T_\rho^*\Z}\langle \xi,u\rangle_{T_\rho\Z}=\iint\!\xi(x,v)\,u(dx\,dv),\\
  T_\rho\Z  &=\M_0(\TT^d\times\RR^d),
  &&
  T^*_\rho\Z=C_b(\TT^d\times\RR^d) \text{ mod constants}.
\end{align*}
For the fluxes $w=(w_\Det,w_\Res)$ we take
\begin{align*}
  \W=T\W &= \M(\TT^d\times \RR^d; \RR^{2d}) \times \M(\TT^d\times \RR^{2d}/2),\\
  T^*_\rho\W&=C_b(\TT^d\times \RR^d; \RR^{2d}) \times C_b(\TT^d\times \RR^{2d}/2) \text{ mod constants},\\
  _{T_\rho^*\W}\langle \zeta,j\rangle_{T_\rho\W}&=\langle\zeta_\Det,j_\Det\rangle+\langle\zeta_\Res,j_\Res\rangle\\
                                                &:=
  \iint_{\TT^d\times\RR^d}\!\zeta_\Det(x,v)\cdot j_\Det(dx\,dv) \\
                                                &\qquad + \iiint_{\TT^d\times\RR^{2d}/2}\!\zeta_\Res(x,v,v')\,j_\Res(dx\,dv\,dv').
\end{align*}
Finally for the continuity operator\footnote{Strictly speaking $\phi$ maps measures to distributions, but we formally restrict to $\Z=\M_1(\TT^d\times\RR^d)$.}:
\begin{align*}
	\phi\lbrack w\rbrack&:= \rho^0-\div w_\Det -\ddiv w_\Res,\\
	d\phi_\rho j&= - \div j_\Det - \ddiv j_\Res, \qquad\qquad
	d\phi_\rho\tp \xi = (\grad \xi,\dgrad\xi),
\end{align*}
where the continuous gradient again acts on the two coordinates $(x,v)$, and the discrete gradient is the adjoint operator of minus the discrete divergence~\eqref{eq:Andersen ddiv}:
\begin{equation*}
  (\dgrad \xi)(x,v,v')=\xi(x,v')-\xi(x,v).
\end{equation*}

\subsection{pre-GENERIC}

At this stage we need to make the specific choice for the resampling measure as a Gaussian
\footnote{The Gaussian is often tweaked by including a nontrivial particle mass $m>0$, and by speeding up or slowing down the resampling by rescaling $\mu$ with a given number $\lambda>0$, see \cite{KLMP2020math}. We omit these factors as they do not significantly change the GENERIC structure.},
and define in addition the measure $\nu$ (assuming $e^{-U}$ is integrable):
\begin{align*}
  \mu(dv)=\frac{e^{-\frac12 \lvert v \rvert^2}}{(2\pi)^{d/2}}, && 	\nu(dx)=\frac{e^{-U(x)}}{\int\!e^{-U(y)}\,dy}.
\end{align*}
The cost $\L$ then indeed satisfies the requirements of Proposition~\ref{prop:preGEN}, and induces the pre-GENERIC structure $(\Psi,\V,b)$ with dissipation potential, free energy and drift (assuming $\rho$ is absolutely continuous),
\begin{align}
  &\Psi^*_\GEN(\rho,\zeta_\Det,\zeta_\Res)\notag\\
  &\qquad :=2\iiint_{\TT^d\times\RR^{2d}/2}\!\sqrt{\rho(x)\mu(v)\rho(x)\mu(v')}\big(\cosh(\zeta_\Res(x,v,v')-1\big)\,dx\,dv\,dv',\notag\\
  &\V(\rho):=\S(\rho\mid \nu\otimes \mu),\qquad\qquad\qquad
  b(\rho):=\begin{bmatrix} j^0_\Det(\rho)\\0\end{bmatrix}.
  \label{eq:Andersen pGEN}
\end{align}
Recall that this means that the large-deviation cost can be decomposed as
\begin{multline*}
  \L(\rho,j)=\Psi_\GEN(\rho,j_\Det-j^0_\Det(\rho),j_\Res)+\Psi^*_\GEN(\rho,-\tfrac12\grad d\V(\rho),-\tfrac12\dgrad d\V(\rho)) \\
  + \langle \tfrac12\grad d\V(\rho),j_\Det\rangle + \langle \tfrac12\dgrad d\V(\rho),j_\Res\rangle,
\end{multline*}
and that the following non-interaction condition holds:
\begin{align*}
  \langle d\phi\tp d\V(\rho),b(\rho)\rangle &= \langle \grad d\V(\rho),j^0_\Det(\rho)\rangle\\
  &=\iint_{\TT^d\times\RR^d}\!\begin{bmatrix} \frac{\grad_x\rho(x,v)}{\rho(x,v)} + \grad_x U(x)\\ \frac{\grad_v\rho(x,v)}{\rho(x,v)} + v\end{bmatrix} 
                        \cdot \begin{bmatrix} v \\ -\grad_x U(x) \end{bmatrix} \,dx\,dv=0.
\end{align*}
In particular, note that $\Psi^*_\GEN(\rho,\zeta_\Det,\zeta_\Res)$ does not depend on $\zeta_\Det$, implying that
\begin{equation}
  \Psi_\GEN(\rho,j_\Det,j_\Res)=\infty \text{ whenever } j_\Det\neq0.
\label{eq:Andersen singularity}
\end{equation}

\subsection{Towards state-space GENERIC and MANERIC} 

The deterministic flux $b(\rho)=(j^0_\Det(\rho),0)$ only becomes Hamiltonian after contracting to the state space:
\begin{align*}
  d\phi_\rho b(\rho)=-\div j^0_\Det(\rho)=-\div_x(\rho v) + \div_v(\rho \grad_x U) = \hat\JJ(\rho)d\E(\rho),
\end{align*}
where
\begin{align}
  \hat\JJ(\rho)\xi&:=-\div\big(\rho(x,v)\begin{bmatrix} 0 &\mathrm{Id}_d \\ -\mathrm{Id}_d & 0 \end{bmatrix} \grad\xi(x,v)\big),\notag\\
  \E(\rho) &:= \iint_{\TT^d\times\RR^d}\!\big(\tfrac12 v^2 + U(x)\big)\,\rho(dx\,dv).  \label{eq:Andersen Hamiltonian}
\end{align}
However, the contracted cost $\hat\L(\rho,u):=\inf_{j:u=-\div j} \L(\rho,j)$ from \eqref{eq:hat L} does \emph{not} induce a (quasi-)GENERIC structure in the state space $\Z$. The reason is that the energy $\E(\rho)$ is not conserved under the full dynamics
\begin{align*}
  \dot\rho(t,x,v)&=d\phi_{\rho(t)}j^0(\rho(t)) \\
  &= -\div\Big(\rho(t,x,v)\begin{bmatrix}v\\-\grad_x U(x)\end{bmatrix}\Big) - \int\!\big\lbrack \rho(x,v)\mu(v')-\mu(v)\rho(x,v')\big\rbrack\,dv'.
\end{align*}
In order to conserve the energy $\E$, one needs to extend the state space as in Proposition~\ref{prop:pGEN to GEN}. In other words, $\hat\L$ does not induce GENERIC, but one can extend the state space to obtain a different cost function that does induce a GENERIC structure. To conclude,
\begin{quote}
  The Andersen thermostat flux cost $\L$ defined in~\eqref{eq:AT cost} \textbf{induces the pre-GENERIC structure} $(\Psi_\GEN,\V,b)$ given by~\eqref{eq:Andersen pGEN}.\\
  Moreover, the contracted drift equation $\dot\rho(t)=d\phi_{\rho(t)} b(\rho(t))$ \textbf{has the Hamiltonian structure} $(\hat\JJ,\E)$ given by~\eqref{eq:Andersen Hamiltonian}.
\end{quote} 
One is then tempted to use Theorem~\ref{th:pGEN to MAN} to extract a MANERIC structure induced by $\L$ from the pre-GENERIC structure. Unfortunately this procedure fails. The reason is that the singularity \eqref{eq:Andersen singularity} causes $\Psi_\GEN$ to be non-differentiable, and so the antisymmetric force~\eqref{eq:forces from pGEN} does not exist and the MFT dissipation potentials $\Psi_\MFT,\Psi^*_\MFT$ in \eqref{eq:PsiMFT from PsiGEN} are nowhere defined.

\section{Example of LDP-induced pre-GENERIC:\\
Zero-range process with deterministic drift}
\label{sec:zero+det}

We aim to provide an example where Theorem~\ref{th:pGEN to MAN} can be used to extract an LDP-induced MANERIC structure from an LDP-induced pre-GENERIC structure. This could not be used on the Andersen thermostat of the previous section due to differentiability issues. The general problem is that on the one hand, one should associate different fluxes to different `directions' so that the non-equilibrium force $F(\rho)$ can be explicitly calculated, but on the other hand, if one of those fluxes is purely deterministic, this will cause the cost function to be non-differentiable, and the force $F(\rho)$ does not exist. At least one deterministic mechanism is however required to induce a (pre-)GENERIC structure.

To provide an example of passing from GENERIC to MANERIC we thus need the deterministic flux to be combined with noise that points in the same direction, so that a singularity of the type~\eqref{eq:Andersen L}, \eqref{eq:Andersen singularity} does not occur. A typical example would be a diffusive system with a suitably chosen deterministic drift, see~\cite[Sec.~5.3]{PattersonRengerSharma2024}. That model is not very instructive because the dissipation potentials are quadratic and so $\Psi^*_\GEN=\Psi^*_\MFT$ by Lemma~\ref{lem:pGEN to MAN}.

In this section we study a model that is perhaps a bit artificial, but it involves a non-quadratic GENERIC dissipation potential $\Psi^*_\GEN$ and satisfies all the requirements needed to pass to MANERIC, resulting in a different MFT dissipation potential $\Psi^*_\MFT$.

\subsection{Microscopic model}

The starting point will be the zero-range process on a graph $\X$ from Section~\ref{sec:zero range} under the same assumptions. The zero range jumps are collected in the cumulative flux $W\super{n}(t)$ as before. This dynamics is now coupled to a deterministic drift $j^0_\Det(\rho)$ (to be determined later in \eqref{eq:zero+det j0}), which is collected in \emph{same} flux $W\super{n}(t)$. This is possible because both mechanisms cause movement in the same direction. We proposed similar models in \cite[Sec.~4.7]{Renger2018a} and \cite[Sec.~3]{Renger2018b}.

The flux is again coupled to the state $\rho\super{n}(t)$ via the continuity equation with discrete divergence~\eqref{eq:discrete divergence},
\begin{equation*}
  \rho\super{n}(t)=-\ddiv W\super{n}(t).
\end{equation*}
Since the deterministic drift causes continuous motion in $\rho\super{n}(t)$, a particle interpretation is no longer appropriate. Rather, the zero-range dynamics causes random jumps of `mass packages' of size $1/n$ from one node to another. The full dynamics of the Markov process $(\rho\super{n}(t),W\super{n})$ is described by the generator:
\begin{multline*}
  (\Q\super{n}f)(\rho,w)=(\Q\super{n}_\ZR f)(\rho,w) \\
                                   \qquad + \grad_\rho f(\rho,w)\cdot(-\ddiv j^0_\Det(\rho)) + \grad_w f(\rho,w)\cdot j^0_\Det(\rho),
\end{multline*}
where $(\Q\super{n}_\ZR f)(\rho,w)$ is the zero-range generator~\eqref{eq:zero range generator}.

\begin{remark} Strictly speaking, the deterministic drift $j^0_\Det(\rho)$ should be suppressed whenever $\rho^0-\ddiv j^0_\Det(\rho)$ hits negative concentrations. For the sake of brevity we ignore this issue.
\end{remark}

\subsection{Macroscopic model, large deviations and geometry}

The macroscopic limit $n\to\infty$ is almost identical to the zero-range case:
\begin{align*}
  \dot \rho_x(t)&=-\ddiv_x \dot w(t), &x\in\X,\\
  \dot w_{xy}(t)&=j^0_{\ZR,xy}\big(\rho(t)\big) + j^0_{\Det,xy}(\rho(t)), &(x,y)\in\X^2/2,
\end{align*}
with zero-range zero-cost flux $j^0_\ZR(\rho)$ from \eqref{eq:zero range limit eqns} and $j^0_\Det(\rho)$ to be determined later.

Similarly, the LDP~\eqref{eq:framework LDP} holds with cost function (recall \eqref{eq:zero range L}):
\begin{align}
  \L(\rho,j)=\L_s\big(\rho,j-j^0_\Det(\rho)\big).
\label{eq:zero+det cost}
\end{align}

The geometry, i.e. the state-flux triple $(\Z,\W,\phi)$, is the same as in the zero-range case, see Subsection~\ref{subsec:zero range geometry}.

\subsection{pre-GENERIC and GENERIC}
\label{subsec:zero+det GEN}

\paragraph{pre-GENERIC.}

Let $\V$ be the quasipotential~\eqref{eq:zero-range QP} for the zero-range process. We now impose additional assumptions on $Q$ and $j^0_\Det$:
\begin{enumerate}[(i)]
\item $\pi_x Q_{xy}=\pi_y Q_{yx}$, i.e. the zero-range dynamics is in detailed balance,
\item $-\ddiv j^0_\Det(\pi)=0$,
\item $\dgrad\grad\V(\rho)\cdot j^0_\Det(\rho)=0$.
\end{enumerate}
Then by construction, $\L$ induces the pre-GENERIC structure $(\Psi_\GEN,\V,j^0_\Det)$, with the dissipation potential from~\eqref{eq:zero-range Psi*}:
\begin{align}
  \Psi^*_\GEN(\rho,\zeta)&=2\sumxly \pi_x Q_{xy} \sqrt{\eta_x(\tfrac{\rho_x}{\pi_x}) \eta_y(\tfrac{\rho_y}{\pi_y}) } \big(\cosh(\zeta_{xy})-1\big),
\label{eq:zero+det Psis}
\end{align}

\paragraph{Hamiltonian structure for the deterministic flow.}

We would now like to choose $j^0_\Det(\rho)$ so that the assumptions above hold, and such that the deterministic flow $\dot\rho(t)=-\ddiv j^0_\Det(\rho(t))$ is a Hamiltonian system. A priori, this flow has as least two conserved quantities: the quasipotential $\V$ due to orthogonality assumption \emph{(iii)} above, and due to the divergence-form of the flow also the total mass:
\begin{align*}
  \E(\rho):=\sum_{x\in\X}\rho_x.
\end{align*}
To simplify we restrict to $\X:=\{1,2,3\}$, and by similar arguments as in Subsection~\ref{eq:zero range Hamiltonian}, the flow must have the form of a cross product between the $\grad\V$ and $\grad\E$, yielding two alternative Hamiltonian structures. Abbreviating $\eta_x=\eta_x(\rho_x/\pi_x)$:
\begin{align}
  -\ddiv j^0_\Det(\rho)&=\grad\E(\rho)\times \grad\V(\rho) 
    = \overbrace{\begin{bmatrix} 0 & -1 & 1 \\ 1 & 0 & -1 \\ -1 & 1 & 0\end{bmatrix}}^{=:\hat\JJ^2(\rho)} \grad\V(\rho)\notag\\
    &= \underbrace{\begin{bmatrix} 0 & \log \eta_3 & -\log\eta_2 & 0 \\ - \log\eta_3 & 0 & \log\eta_1 \\ \log \eta_2 & -\log\eta_1 & 0\end{bmatrix}}_{=:\hat\JJ(\rho)} \grad\E(\rho)
    = \begin{bmatrix} \log \eta_3/\eta_2 \\ \log \eta_1/\eta_3 \\ \log \eta_2/\eta_1 \end{bmatrix}.
\label{eq:zero+det Ham} 
\end{align}
Observe that Assumption~\emph{(ii)} is indeed satisfied.

A possible choice for the deterministic flux $j^0_\Det(\rho)$ giving rise to this Hamiltonian system in state space would be:
\begin{equation}
  j^0_{\Det,xy}(\rho) := \log\big(\eta_x(\rho_x/\pi_x) \eta_y(\rho_y/\pi_y)\big), \qquad (x,y)\in\X^2/2.
\label{eq:zero+det j0}
\end{equation}
As far as the author is aware, it is not possible to find a $j^0_\Det(\rho)$ satisfying \eqref{eq:zero+det Ham} such that $\dot w(t)=j^0_\Det(\rho^0-\ddiv w(t))$ is Hamiltonian, which is, as argued throughout this paper, not to be expected.

\begin{remark}
  Since $\V,\E$ are conserved along the deterministic flow, any functions thereof $f\circ\V, g\circ\E$ will also be conserved. This gives extra flexibility in chosing the Hamiltonian structure.
\end{remark}

\paragraph{GENERIC in state space.}

Under the choice \eqref{eq:zero+det j0}, the cost $\L(\rho,j)$ does not induce GENERIC, but the contracted cost $\hat\L(\rho,u):=\inf_{j:u=-\div j} \L(\rho,j)$ does, with the contracted dissipation potentials (see \cite{Renger2018b}):
\begin{align}
  \hat\Psi_\GEN(\rho,u):=\inf_{j\in T_\rho\W:-\ddiv j = u} \Psi_\GEN(\rho,j), &&
  \hat\Psi^*_\GEN(\rho,\xi)=\Psi^*_\GEN(\rho,\dgrad\xi).
\label{eq:zero+det contracted diss}
\end{align}

Recall the definition of $\hat\L$-induced GENERIC from Definition~\ref{def:rho GEN}. The conservation of $\V$ (that is, assumption \emph{(iii)}) under the deterministic flow is exactly one of the non-interaction conditions of GENERIC. In order to satisfy the other non-interaction condition, we must chose the Hamiltonian structure $\hat\JJ,\E$. Indeed,
\begin{align*}
  \hat\Psi^*_\GEN\big(\rho,\xi+\theta d\E(\rho)\big) = \Psi^*_\GEN\big(\rho,\dgrad\xi+\theta\underbrace{\dgrad\grad\E(\rho)}_{=0}\big)= \hat\Psi^*_\GEN(\rho,\xi).
\end{align*}
This argument would break down if we were to chose the other Hamiltonian structure $(\hat\JJ^2,\V)$. We conclude that
\begin{quote}
  With the choice~\eqref{eq:zero+det j0} for $j^0_\Det$, the cost function $\L$ defined in~\eqref{eq:zero+det cost} \textbf{induces the pre-GENERIC structure} $(\Psi_\GEN,\V,j^0_\Det)$ with dual dissipation potential $\Psi^*_\GEN$ and free energy $\V$ given by \eqref{eq:zero+det Psis} and ~\eqref{eq:zero-range QP}. \\
  The corresponding contracted cost function $\hat\L$ \textbf{induces the GENERIC structure} $(\hat\Psi_\GEN,\V,\hat\JJ,\E)$ with contracted dissipation potential $\hat\Psi_\GEN$, quasipotential $\V$ and Hamiltonian structure $(\hat\JJ,\E)$ given by \eqref{eq:zero+det contracted diss}, \eqref{eq:zero-range QP}, and~\eqref{eq:zero+det Ham}.
\end{quote}

\begin{remark} The two Poisson structures $\hat\JJ,\hat\JJ^2$ satisfy the Jacobi identity as a property of the cross product. The structure $(\hat\JJ^2,\V)$ can be trivially generalised to arbitrary dimensions; for $(\hat\JJ,\E)$ this is not trivial.
\end{remark}

\subsection{From GENERIC to MANERIC}

The results of the previous subsection are sufficient condition for the application of Theorem~\ref{th:pGEN to MAN}. Note in particular that the dissipation potentials~\eqref{eq:zero+det contracted diss}, \eqref{eq:zero-range Psi*} are even and so the free energy $\V$ that drives the detailed balance part of the dynamics is also the quasipotential in the MFT sense (i.e. the Hamilton-Jacobi equation~\eqref{eq:HJE} holds). 

\paragraph{Force structure.}
The MFT forces and dissipation potential can be explicitly calculated using Lemma~\ref{lem:pGEN to MAN} (or Proposition~\ref{prop:force structure}). This yields, again abbreviating $\eta_x:=\eta_x(\rho_x/\pi_x)$:
\begin{align}
  F^\sym_{xy}(\rho)  &=-\tfrac12\dgrad_{xy}\grad\V(\rho)=\frac12 \log\frac{\eta_x}{\eta_y},         \notag\\
  F^\asym_{xy}(\rho) &={\cosh^*}'\big( \mfrac{ j^0_{\Det,xy}(\rho)}{2\sqrt{\pi_x Q_{xy}\pi_y Q_{yx}\eta_x\eta_y}}\big)
                      =\asinh\big(\mfrac{\log \eta_x\eta_y}{2\pi_x Q_{xy}\sqrt{\eta_x\eta_y}}\big), \label{eq:zero+det Fasym}\\
  \Psi^*_\MFT(\rho,\zeta)  &=\sumxly \sqrt{(\log(\eta_x\eta_y))^2+(2\pi_x Q_{xy})^2 \eta_x\eta_y}\big(\cosh(\zeta_{xy})-1\big) \notag\\[-0.7em]
                     &\hspace{16em} + \big(\zeta_{xy}-\sinh(\zeta_{xy})\big)\log(\eta_x\eta_y).
  \label{eq:zero+det PsisMFT}
\end{align}
Setting $(2\pi_x Q_{xy})^2 \eta_x\eta_y\geq0$ shows that indeed $\Psi^*_\MFT(\rho,\zeta)\geq0$, and moreover $\Psi^*_\MFT(\rho,0)=0$, so that $\Psi_\MFT$ is indeed a dissipation potential. However, although $\Psi_\GEN$ is even, this is no longer the case for $\Psi_\MFT$.

\paragraph{MANERIC.}

By construction,
\begin{align*}
  -\ddiv \grad_\zeta\Psi^*(\rho,F^\asym(\rho))=-\ddiv j^0_\Det(\rho)=\hat\JJ(\rho)\grad\E(\rho)=\hat\JJ^2(\rho)\grad\V(\rho),
\end{align*}
with the two alternative Hamiltonian structures $(\hat\JJ,\E),(\hat\JJ^2,\V)$ from~\eqref{eq:zero+det Ham}. We conclude that, in addition to the (pre-)GENERIC structures of Subsection~\ref{subsec:zero+det GEN},
\begin{quote}
  The cost function $\L$ defined in~\eqref{eq:zero+det cost} \textbf{induces the force structure} $(\Psi_\MFT,F)$ with (non-even) dissipation potential $\Psi^*_\MFT$ given by~\eqref{eq:zero+det PsisMFT} and force $F=F^\sym+F^\asym$ with symmetric and antisymmetric forces~\eqref{eq:zero+det Fasym}.\\
  Furthermore, $\L$ \textbf{induces the MANERIC structures} $(\Psi_\MFT,F,\V,\hat\JJ,\E)$ and $(\Psi_\MFT,F,\V,\hat\JJ^2,\V)$, where the two Hamiltonian structures are given by~\eqref{eq:zero+det Ham}.
\end{quote}



\section*{Acknowledgements}

The author thanks Upanshu Sharma (UNSW Sydney) and Michal Pavelka (Charles University Prague) for the useful discussions.

\appendix

\section{Gradient flows and Hamiltonian systems}
\label{sec:GFs and HSs}

Classic gradient flows are evolution equations of the form (written in flux space, and assuming the geometry depends on $w$ through $\rho$ only)
\begin{equation*}
  \dot w(t)=-\Grad_{w(t)}\S(w(t))=-\KK(\rho(t)) d\S(w(t)), 
\end{equation*}
where $\S:\W\to\RR$ is a differentiable functional and $\KK(w):T_w^*\W\to T_w\W$ a positive semidefinite linear ``Onsager operator'' (the inverse of the metric tensor). The definiteness guarantees that along the flow the second law of thermodynamics is uphold: $d/dt\,\S(w(t))=-\langle d\S(w(t),\KK(w(t))d\S(w(t))\rangle\leq0$. The equivalent energetic formulation of a gradient flow, called the \underline{E}nergy-\underline{D}issipation-\underline{I}nequality reads:
\begin{align*}
  &\tfrac12\lVert\dot w(t)\rVert^2_{\KK(w(t))^{-1}} + \tfrac12\lVert d\S(w(t))\rVert^2_{\KK(w(t))} +\langle d\S(w(t)),\dot w(t))\rangle =0, \quad\text{with}\\
  &\tfrac12\lVert j\rVert^2_{\KK(w)^{-1}}:=\sup_{\zeta\in T_w\W}\langle \zeta,j\rangle - \tfrac12\lVert\zeta\rVert^2_{\KK(w)}
  \quad\text{and}\quad
  \tfrac12\lVert \zeta\rVert^2_{\KK(w)}:=\tfrac12\langle\zeta,\KK(w)\zeta\rangle.
\end{align*}
The two squared norms are ``quadratic dissipation potentials''.

Many applications require the use of convex, not necessarily quadratic dissipation potentials. The origins of these can be traced back to the work of Marcelin~\cite{Marcelin1915}, \cite{ColVis90CDNE}, \cite{Luckhaus1995}; applications to statistical theories and fully non-detailed balanced systems are more recent, see for example \cite{RossiMielkeSavare08,MielkePeletierRenger2014,KaiserJackZimmer2018,PattersonRengerSharma2024}. This generalises to gradient flow equations of the form
\begin{align*}
  &\dot w(t)=d_\zeta\Psi^*\big(w(t),-d\S(w(t))\big), \qquad\qquad\text{or equivalently}\\
  &\Psi\big(w(t),\dot w(t)\big) + \Psi^*\big(w(t),-d\S(w(t))\big) +\langle d\S(w(t)),\dot w(t))\rangle =0,
\end{align*}
where $\Psi,\Psi^*$ satisfy the following.
\begin{definition} $\Psi:T\W\to\RR\cup\{\infty\}$ and $\Psi^*:T^*\W\to\RR\cup\{\infty\}$ are called \emph{convex dual dissipation potentials} whenever
\begin{enumerate}[(i)]
  \item $\Psi^*(\rho,\zeta)=\sup_{j\in T_\rho\W}\langle\zeta,j\rangle - \Psi(\rho,j)$,
  \item $\Psi(\rho,j)=\sup_{\zeta\in T_\rho\W^*}\langle\zeta,j\rangle - \Psi^*(\rho,\zeta)$, 
  \item $\inf_{j\in T_\rho\W}\Psi(\rho,j)=\Psi^*(\rho,0)=0$ and $\inf_{\zeta\in T^*_\rho\W}\Psi^*(\rho,\zeta)=\Psi(\rho,0)=0$.
\end{enumerate}
\label{def:disspots}
\end{definition}
These resulting (generalised) gradient flows are also thermodynamically sound due to the non-negativity of both dissipation potentials: $d/dt\,\S(w(t))=\langle d\S(w(t)),\dot w(t))\rangle=-\Psi\big(w(t),\dot w(t)\big) - \Psi^*\big(w(t),-d\S(w(t))\big)\leq0$.

We now recall the definition of a Hamiltonian system. 
\begin{definition}\phantom{a}
\begin{itemize}
\item
A linear operator $\JJ(\rho):T_\rho^*\Z\to T_\rho\Z$ , differentiable in $\rho\in\Z$, is called a \emph{Poisson structure} if it satisfies the Jacobi identity (and skewsymmetry as a consequence), that is for all twice differentiable functionals $F,G,H:\Z\to\RR$,
\begin{multline*}
  \big\langle d F(\rho),\JJ(\rho)d\langle dG(\rho),\JJ(\rho)dH(\rho)\rangle\big\rangle
  +
  \big\langle d G(\rho),\JJ(\rho)d\langle dH(\rho),\JJ(\rho)dF(\rho)\rangle\big\rangle\\
  +
  \big\langle d H(\rho),\JJ(\rho)d\langle dF(\rho),\JJ(\rho)dG(\rho)\rangle\big\rangle  
  \equiv0.
\end{multline*}
Equivalently, the bracket $\lbrack F,G\rbrack:= \langle dF(\rho),\JJ(\rho) dG(\rho)\rangle$ is a Lie group.
\label{def:Poisson structure}
\item An evolution equation $\dot\rho(t)=\A(\rho(t))$ for some operator $\Z\ni\rho\mapsto\A(\rho)\in T_\rho\Z$ has the Hamiltonian structure $(\JJ,\E)$ for some Poisson structure $\JJ$ and (G{\^a}teaux) differentiable functional $\E:\Z\to\RR$ if
\begin{equation*}
  \A(\rho)=\JJ(\rho)d\E(\rho).
\end{equation*}
\end{itemize}
\label{def:Ham structure}
\end{definition}

\section{Hamiltonian structures for arbitrary evolutions}
\label{app:any system Hamiltonian}

In this appendix we show that \emph{any} evolution equation $\dot\rho(t)=b(\rho(t))$ can be supplemented with additional variables $\tilde\rho(t)=(\rho(t),\cdot)$ so that the extended evolution equation $\dot{\tilde\rho}(t)=\tilde b(\tilde\rho(t))$ has a Hamiltonian structure. As a consequence, saying that an evolution equation has a Hamiltonian structure \emph{after} extending the variables is in itself an empty statement, and says nothing about the behaviour of the original system. 

Although this result may be surprising at first sight, we provide two alternative constructions that are both rather trivial.

\begin{proposition}[Construction \#1, taken from \cite{KLMP2020math}] Consider an evolution equation $\dot\rho(t)=b(\rho(t))$ with operator $b:\Z\to T\Z$, and define the extended variables
$\tilde\rho(t):=(\rho(t),e(t))$, where $e(t)\equiv e(0)\in\RR$ is a constant, real-valued energy. Define
\begin{align*}
\E(\rho,e):=e
&&
\JJ(\rho,e):=
  \begin{bmatrix}
    0 & b(\rho)\cdot \\
    -\langle \cdot,b(\rho)\rangle & 0
  \end{bmatrix},
\end{align*}
where $b(\rho)\cdot$ means scalar multiplication and $\langle \cdot,b(\rho)\rangle$ means the canonical dual pairing between $T_\rho^*\Z$ and $T_\rho\Z$. Then
\begin{align*}
  \dot{\tilde \rho}(t)=\begin{bmatrix}\dot\rho(t)\\ \dot e(t)\end{bmatrix} = \JJ\big(\rho(t),e(t)\big)\,d\E\big(\rho(t),e(t)\big)=\begin{bmatrix} b\big(\rho(t)\big)\\0\end{bmatrix},
\end{align*}
and the operator $\JJ(\rho,e)$ is a Poisson structure (i.e. is skew-symmetric and satisfies the Jacobi identity)~\cite[Th.~4.1 and Prop.~A.1]{KLMP2020math}.
\label{prop:Ham struct 1}
\end{proposition}

Note that for this construction the choice of the initial energy $e(0)$ has no effect on the dynamics of $\rho(t)$.

\begin{proposition}[Construction \#2, classic, see for example~\cite{EsenGrmelaPavelka2022} or {\cite[Th.~3.3.2]{Evans2010}}] Consider an evolution equation $\dot\rho(t)=b(\rho(t))$ with operator $b:\Z\to T\Z$, and a Lagrangian cost function $L:T\W\to\RR$ so that $L(\rho,u)$ is its own differentiable and strictly convex bidual in $v$ and $\inf_{v\in T_\rho\Z} L(\rho,u)=L\big(\rho,b(\rho)\big)$. 

Define $\tilde\rho(t)=(\rho(t),p(t))$ with `momentum' $p(t):=d_v L\big(\rho(t),v(t)\big)\in T_{\rho(t)}^*\Z$, and let $H(\rho,\xi):=\sup_{v\in T_\rho\Z}\langle\xi,v\rangle- L(\rho,u)$. Then
\begin{align*}
  \dot{\tilde \rho}(t)=\begin{bmatrix}\dot\rho(t)\\ \dot p(t)\end{bmatrix} = \begin{bmatrix} 0 & I \\ -I & 0\end{bmatrix}dH\big(\rho(t),p(t)\big),
\end{align*}
with the simplectic Poisson structure $\JJ(\rho,p):=([0\,I],[-I\,0])$ and conserved energy $H(\rho(t),p(t))$.
\label{prop:Ham struct 2}
\end{proposition}

\begin{proof}
A trajectory $\rho(\cdot)$ satisfies $\dot\rho(t)=b(\rho(t))$ for almost all $t\in[0,T]$ if and only if it minimises the action functional $\rho\mapsto\int_0^T\!L\big(\rho(t),\dot\rho(t)\big)\,dt$, which implies the corresponding Euler-Lagrange equation:
\begin{align*}
  \dot p(t)=\frac{d}{dt} d_v L\big(\rho(t),\dot\rho(t)\big)=d_\rho L\big(\rho(t),\dot\rho(t)\big)=-d_\rho H(\rho(t),p(t)).
\end{align*}
On the other hand $\dot\rho(t)=d_v H(\rho(t),p(t))$ by convex duality and the definition of $p$, showing that Hamilton's equations hold.
\end{proof}

Note that for the second construction, the Euler-Lagrange equation introduces new solutions to the Lagrangian minimisation problem, indexed by the initial momentum $p(0)$. The original dynamics $\rho(t)=b(\rho(t))$ is retrieved by choosing the $p(0):=0$. Indeed by assumption $\dot p(0)=d_\rho H(\rho(0),0)=d_\rho\big(-\inf_v L(\rho,u)\big)\equiv0$, and so $p(t)\equiv 0$. In that case the conserved energy will simply be $H(\rho(t),p(t))=H(\rho(t),0)\equiv0$, and $\dot\rho(t)=d_\xi H(\rho(t),0)=b(\rho(t))$.

We also like to stress that for a given evolution equation, this construction works for \emph{any choice} of the cost $L$. In the context of our paper, a natural choice would be $L(\rho,u):=\hat\Psi^*(\rho,v-b(\rho))$, if a dual dissipation potential $\hat\Psi^*:T^*\Z\to\RR$ is available.

\section{A little bit of convex analysis}

To simplify notation and to stress the generality we work in the setting of a Banach space $\X$ and its dual $\X^*$.

\begin{proposition}[{\cite[Prop.~2.1]{MielkePeletierRenger2014}}] Let $l:\X\to\lbrack0,\infty\rbrack$ be differentiable, convex and lower semicontinuous (and so $l$ is its own convex bidual $h:\X^*\to(-\infty,\infty\rbrack$), and assume $\inf l=0$. Then there exist unique convex dual dissipation potentials $\psi:\X\to\lbrack0,\infty\rbrack, \psi^*:\X\to\lbrack0,\infty\rbrack$ and $f^*\in \X^*$ such that
\begin{equation}
  l(x) \equiv \psi(x) + \psi^*(f^*) - \langle f^*,x\rangle.
\label{eq:app force structure}
\end{equation}
These are given by
\begin{align*}
  f^*=-dl(0), && \psi^*(x^*)=h(x^*-f^*)-h(-f^*), && \psi(x)=\sup_{x\in\X} \langle x^*,x\rangle-\psi^*(x^*).
\end{align*}
Moreover, $l(x)\equiv(-x)-2\langle x^*,x\rangle$ if and only $\psi$ is even if and and only if $\psi^*$ is even.
\label{prop:MPR}
\end{proposition}
\begin{proof} Recall from Definition~\ref{def:disspots} that $d\psi(0)=0$ and $\psi^*(0)=0$.

The existence, uniqueness and explicit expression of $f^*$ is then obtained $-d l(0)=-d\psi(0)+f^*=f^*$.

Claim~\eqref{eq:app force structure} with unique dissipation potentials is equivalent to $h(x^*)\equiv \psi^*(x^*+f*)-\psi^*(f^*)$, and so, choosing $y^*=f^*$,
\begin{align*}
  \psi^*(x^*)&=h(x^*-f^*)+\psi^*(f^*)=h(x^*-f^*)-h(y^*)+\psi^*(y^*+f^*)\\
   &=h(x^*-f^*)-h(-f^*)+\psi^*(0)=h(x^*-f^*)-h(-f^*).
\end{align*}
The evenness of the dissipation potentials becomes trivial after inserting \eqref{eq:app force structure} in the symmetry relation.
\end{proof}

\begin{proposition} Let $l:\X\to\lbrack0,\infty\rbrack$ be convex, and assume there exist $y\in\X,f^*\in\X^*$ and convex dual $\psi:\X\to(-\infty,\infty\rbrack, \psi^*:\X^*\to(-\infty,\infty\rbrack$ so that
\begin{equation*}
  l(x) \equiv \psi(x+y) + \psi^*(f^*) - \langle f^*,x\rangle.
\end{equation*}
If $\inf l=0$ then $\langle f^*,y\rangle\equiv0$.
\label{prop:incomplete L}
\end{proposition}
\begin{proof}
\begin{align*}
  0=-\inf l &= \sup_{x\in\X} \langle f^*,x\rangle-\psi(x+y)-\psi^*(f^*) \\&= \sup_{x\in\X} \langle f^*,x-y\rangle-\psi(x)-\psi^*(f^*)=\psi^*(f^*)-\psi^*(f^*)-\langle f^*,y\rangle.
\end{align*}
\end{proof}

\section{Quasipotential for the zero-range $\MM$-structure}
\label{sec:M-structure}

This section is dedicated to the cost function~\eqref{eq:quadratised L M} derived from the original large-deviation cost function $\L_s$ of the zero-range process:
\begin{align*}
  \L_\MM(\rho,j)&:=\mfrac12\lVert j-j^0(\rho)\rVert^2_{\MM(\rho)^{-1}}:=\sup_{\zeta\in T_\rho^*\W} \langle\zeta,j\rangle - \H_\MM(\rho,\zeta),\\
  \H_\MM(\rho,\zeta)&:=\mfrac12 \langle \zeta,\MM(\rho)\zeta\rangle + \langle \zeta,j^0(\rho)\rangle,
\end{align*}
the Onsager operator $\MM$ being given by \eqref{eq:zero-range MM}. In order to apply the MFT arguments of Section~\ref{sec:MFT}, one needs to
find the corresponding quasipotential $\V_\MM$ that solves the Hamilton-Jacobi equation~\eqref{eq:HJE}, for almost all $\rho\in\Z$:
\begin{align*}
  \H_\MM\big(\rho,\dgrad\grad\V_\MM(\rho)\big)=0, && \inf_{\rho\in\Z}\V_\MM(\rho)=0.
\end{align*}
This is a non-trivial problem, and we only present a number of properties.

The first property says that the detailed balance condition is the same as detailed balance for the Markov chain $Q$.
\begin{proposition} If $\pi_x Q_{xy}=\pi_y Q_{yx}$ for all $x,y\in\X$, then:
\begin{enumerate}[(i)]
\item $\V_{\MM}(\rho)=\V(\rho)$ from~\eqref{eq:zero-range QP},
\item $F(\rho)=F^\sym(\rho)=-\tfrac12\dgrad\grad\V(\rho)$ and hence $F^\asym(\rho)\equiv0$.
\end{enumerate}
\label{prop:white noise DB}
\end{proposition}
\begin{proof} \emph{(i)} Abbreviate $\eta_x=\eta_x(\rho_x/\pi_x)$. Since $\pi_x Q_{xy}=\pi_y Q_{yx}$ the logarithmic mean can be rewritten as (coinciding with \cite{Mielke2011GF},\cite{Maas2011}),
\begin{equation*}
  \Lambda(\pi_x Q_{xy}\eta_x,\pi_yQ_{yx}\eta_y)=\pi_x Q_{xy}\Lambda(\eta_x,\eta_y).
\end{equation*}
Using $\dgrad_{xy}\grad\V(\rho)=\log\frac{\eta_y}{\eta_x}$, we verify
\begin{align*}
  \H_\MM\big(\rho,\dgrad\grad\V(\rho)\big)
  &=\sumxly \big( \log\mfrac{\eta_y}{\eta_x} \big)^2 \pi_x Q_{xy} \Lambda(\eta_x,\eta_y) \\
  &\hspace{1em} + \sumxly \log\mfrac{\eta_y}{\eta_x} ( \pi_xQ_{xy}\eta_x-\pi_y Q_{yx}\eta_y)=0.
\end{align*}
\emph{(ii)} By construction, $\L_\MM$ induces the force structure $(\Psi_\MM,F)$ with force~\eqref{eq:zero-range force}, and so:
\begin{equation*}
  F_{xy}(\rho)=\frac12\log\frac{\pi_x Q_{xy} \eta_x}{\pi_y Q_{yx} \eta_y}=-\frac12\log\frac{\eta_y}{\eta_x}=-\mfrac12\dgrad_{xy}\grad\V(\rho).
\end{equation*}
\end{proof}
This result does not necessarily mean that the quasipotential is also $\V_\MM=\V$ in the non-detailed balanced setting. In fact, we can show that it is not.

\begin{proposition}
Assume that $\pi_x Q_{xy}\neq\pi_y Q_{yx}$ for some $x,y\in\X$. Then $\H_\MM\big(\rho,\dgrad\grad\V(\rho)\big)\neq0$ on a non-null set of points $\rho\in\P(\X)$.
\label{prop:zero-range no QP}
\end{proposition}
\begin{proof}
Let
\begin{equation*}
  (0,1)\ni\alpha\mapsto\rho_z^\alpha:=
  \begin{cases}
    \alpha\pi_x, &z=x,\\
    (1-\alpha)\pi_y, &z=y,\\
    \pi_z, &z\neq x,y,
  \end{cases}
\end{equation*}
and abbreviate again $\eta_x^\alpha:=\eta_x(\rho^\alpha_x/\pi_x)$. We further introduce the following function:
\begin{equation*}
  \theta(\beta;f,g):=\big(\log(\beta)\big)^2\Lambda(f,g)+\big(\log(\beta)\big)\cdot(f-g),
\end{equation*}
so that
\begin{align*}
  \lim_{\beta\to0}\theta(\beta;\, f,\beta g) = -f\log f/g,
  &&
  \lim_{\beta\to0}\theta\big(1-\beta,f,(1-\beta) g\big) = 0.
\end{align*}
Using the symmetry of $\Lambda(\cdot,\cdot)$ and the inversion rule $\theta(\beta;f,g)=\theta\big(\tfrac{1}{\beta};g,f\big)$,
\begin{align*}
  &\H_\MM\big(\rho^\alpha,\dgrad\grad\V(\rho^\alpha\big)
  =\theta\big( \tfrac{\eta^\alpha_x}{\eta^\alpha_y};\, \pi_y Q_{yx}\eta^\alpha_y,\pi_x Q _{xy}\eta^\alpha_x\big)\\
  &\hspace{5em}+\sum_{\substack{z\in\X:\\z\neq x, y}} \Big\lbrack\theta(\eta^\alpha_x;\, \pi_z Q_{zx}\eta^\alpha_z,\pi_x Q_{xz}\eta^\alpha_x)
  + \theta(\eta^\alpha_y;\, \pi_z Q_{zy}\eta^\alpha_z,\pi_y Q_{yz}\eta^\alpha_y) \, \Big\rbrack,
  \\&\qquad\xrightarrow{\alpha\to0} - \sum_{\substack{z\in\X:\\z\neq x}} \pi_z Q_{zx}\log\big(\mfrac{\pi_z Q_{xz}}{\pi_x Q_{xz}}\big),
\end{align*}
where the remaining terms drop out because $\dgrad_{ab}\grad\V(\rho^\alpha)=0$ for all $a<b$ with $a,b\neq x,y$.

The right-hand side is nonpositive by Jensen's inequality, and equality holds only if $\pi_{z}Q_{zx}/(\pi_x Q_{xz})=:\kappa$ is constant in $z$. However, $\kappa\neq1$ by assumption ($y$ is included in the last sum), and if $\kappa>1$ or $\kappa<1$ then in the steady state more mass would flow into than out of $x$ or vice versa, which contradicts $\pi$ being the steady state.

It follows that $\H_\MM\big(\rho(\alpha),\dgrad\grad\V(\rho(\alpha)\big)<0$ for all $\alpha>0$ sufficiently small, which proves the claim.
\end{proof}

To summarize, it is unclear what the quasipotential is for the cost $\L_\MM$, and as a result many expressions of the theory developped in Section~\ref{sec:MFT} can not be made explicit. As described in \cite[Sec.~IV.E]{BDSGJLL2015MFT} for lattice gases, we expect the quasipotential to be a non-local function, due to the occurrence of long-range correlations.



\bibliographystyle{alpha}
\bibliography{bib} 

\end{document}